%% file: 0_Main_new_format.tex
\documentclass[final,3p,times,twocolumn]{elsarticle}
\usepackage{amsmath}
\usepackage{amsfonts}
\usepackage{amssymb}
\usepackage{amsthm}
\usepackage{color}
\usepackage{multirow}
\usepackage{algorithm}
\usepackage{algpseudocode}
\usepackage{xspace}
\usepackage{makecell}
\usepackage{pifont}
\usepackage{color}
\usepackage{amsmath}
\usepackage{dirtytalk}
\usepackage{amssymb}
\usepackage{amsthm}
\usepackage{epstopdf}
\usepackage{caption}
\usepackage{subcaption}
\usepackage{mathtools}
\mathtoolsset{showonlyrefs}
\usepackage{appendix}
\usepackage{tabularx,colortbl}
\usepackage{tikz}
\usepackage{pgfplots}
\usetikzlibrary{patterns, shapes, arrows, calc}
\pgfplotsset{compat=newest}
\usepackage[nice]{nicefrac}

\newtheorem{assumption}{Assumption}
\theoremstyle{definition}
\newtheorem{definition}{Definition}
\newtheorem{exmp}{Example}
\newtheorem{remark}{Remark}

\usepackage[most]{tcolorbox}

\newtcolorbox{takehome}[1]
{
	colback=red!3!white,
	colframe=red!75!white,  
	title={#1},
}
\usepackage{graphicx}
\usepackage[colorlinks=true, allcolors=blue]{hyperref}

\journal{Annual Reviews in Control}

\begin{document}

\begin{frontmatter}

\title{Quantifying Security for Networked Control Systems: A Review \tnoteref{t1}}
\tnotetext[t1]{This research is supported by the Swedish Research Council grants 2024-00185, 2023-05234, 2021-06316, by the Swedish Foundation for Strategic Research, and by the Knut and Alice Wallenberg Foundation.}

\author[KTH]{Sribalaji C. Anand}
\ead{srca@kth.se}
\author[UU]{Anh Tung Nguyen}
\ead{anh.tung.nguyen@it.uu.se}
\author[UU]{Andr\'e M.H. Teixeira }
\ead{andre.teixeira@it.uu.se}
\author[KTH]{Henrik Sandberg}
\ead{hsan@kth.se}
\author[KTH]{Karl H. Johansson}
\ead{kallej@kth.se}

\affiliation[KTH]{organization={Division of Decision and Control Systems, KTH Royal Institute of Technology, and Digital Futures}, city={Stockholm}, country={Sweden}}
\affiliation[UU]{organization={Division of Systems and Control, Department of Information Technology, Uppsala University}, city={Uppsala}, country={Sweden}}
%
%

\begin{abstract}
Networked Control Systems (NCSs) are integral in critical infrastructures such as power grids, transportation networks, and production systems. Ensuring the resilient operation of these large-scale NCSs against cyber-attacks is crucial for societal well-being. Over the past two decades, extensive research has been focused on developing metrics to quantify the vulnerabilities of NCSs against attacks. Once the vulnerabilities are quantified, mitigation strategies can be employed to enhance system resilience. This article provides a comprehensive overview of methods developed for assessing NCS vulnerabilities and the corresponding mitigation strategies. Furthermore, we emphasize the importance of probabilistic risk metrics to model vulnerabilities under adversaries with imperfect process knowledge. The article concludes by outlining promising directions for future research.
\end{abstract}
%
%
\end{frontmatter}
\tableofcontents
\section{Introduction}
\input{1_Intro.tex}
\section{Problem Setup}\label{sec:problem}
\input{Problem.tex}
\section{Impact metrics}\label{sec:impact}
In this section, we review articles 
{that} aim to solve the optimization problem \eqref{eq:impact}.
\input{Impact.tex}
\section{Resource metrics}\label{sec:security}
In this section, we review articles that quantify the resources needed by the adversary to conduct a stealthy attack \eqref{eq:security}. As discussed in Example~\ref{exmp:resource}, the resources relate to the likelihood of an attack. Similar to computer security \cite{ben2013using}, if the resources needed by the adversary are low (high), then the likelihood of the attack scenario can be presumed high (low). 
\input{Resource}
\section{Risk metrics}\label{sec:risk}
\input{Risk.tex}
\section{Metrics for specific applications}\label{sec:application}
In this section, we review the works that develop security metrics for power grids. We also review applications used as numerical examples in the security literature to depict the effectiveness of attack policies.
\input{Application.tex}
\section{Attack mitigation}\label{sec:mitigation}
\input{Mitigate.tex}
\section{Outlook}\label{sec:con}
This survey provided a comprehensive overview of the current landscape of the security literature. However, several topics lie beyond the scope of this review. For instance, security can be quantified the detector output such as True Positive detection Rate {(TPR)} and False Positive detection Rate (FPR). In this article, we did not focus on these metrics but rather on quantifying the physical damage caused by the adversary. Readers interested in quantifying security using FPR and TPR are referred to the work \cite[Table 1]{urbina2016limiting}. 

Similarly, some mitigation strategies, such as the multi-observer approach \cite{yang2018multi}, are not stated here since they do not consider the impact inflicted on the closed-loop system or the resources needed for the adversary. In the remainder of this section, we highlight several promising avenues where further research investigation could yield valuable insights.
\input{Future.tex}
\section{Conclusions}
This review article explored the area of quantifying security in NCSs. Unlike traditional computer security frameworks, which often neglect the physical consequences of attacks, we focus on the analysis and mitigation of physical degradation caused by attacks. We divided the existing literature into two broad categories: impact metrics which quantify the physical degradation a stealthy attack, and resource metrics which is related to the likelihood of an attack scenario.

We also highlighted the metrics developed for large-scale multi-agent systems and power grids. Finally, we discussed how the uncertainty of the operator and/or the adversary about the process knowledge can be built into the impact metrics using probabilistic risk metrics. Together, these metrics serve as tools for designing resilient and proactive mitigation strategies. We also discussed the mitigation strategies developed for each metric in the literature.

Despite substantial progress, several open challenges remain. Existing methods often rely on idealized assumptions such as complete knowledge of the system dynamics. Moreover, computational tractability remains a bottleneck for many real-world applications. Additionally, integrating these quantification approaches into system design tools could further advance the goal of \emph{secure-by-design} NCSs.

In summary, this review underscores that quantification of NCS security is not merely an academic exercise but a necessary step toward designing resilient networked control systems. As NCSs continue to form the backbone of critical infrastructure, developing robust methods for quantifying and mitigating cyber risks remains a pressing area of research.
\bibliographystyle{elsarticle-num}
\addcontentsline{toc}{section}{References}
\bibliography{mybref}
\end{document}

%% file: 1_Intro.tex
On a daily basis, traffic lights coordinate to reduce congestion \citep{chen2010review}, power grids balance energy demands in real-time \citep{wang2011wide}, and water distribution systems monitor and adjust flow to meet changing needs \citep{amin2012cyber}. Since these systems are essential for the nation's economy, public health, and safety, they are termed \emph{Critical infrastructures}.
These critical infrastructures are effectively managed by a group of automated decision-making mechanisms. The combination of the decision-making mechanisms and physical processes is called a \emph{control system} that keeps critical infrastructure operating smoothly. 

As technology evolves and human demands increase, control systems are becoming more complex, intelligent, and interconnected (often through wireless networks and the internet). Equipped with these interconnections, control systems evolve into networked control systems (NCSs). However, due to these interconnections, NCSs also open the door to cyber threats that were unfortunately unimaginable when these new technologies were first designed. These cyber threats, deployed by malicious adversaries, lead not only to eavesdropping on confidential information in NCSs but also to catastrophic physical damages \cite{hemsley2018history}. 

Proactive actions against these adversaries are obviously demanded by the defender, which leads to research on security in NCSs. Security in NCSs is a societal issue rather than just a technical one. A malicious attack on a power grid can cause blackouts that disrupt lives, economies, and even healthcare \cite{slowik2019crashoverride}. An intentional discharge of sewage from a water treatment system can poison a large living area of creatures and civilians \cite{slay2007lessons}. These are not science fiction scenarios since they have already happened in Ukraine \cite{slowik2019crashoverride}, Australia \cite{slay2007lessons}, Europe \cite{Havex2014}, and Israel \cite{Israel2020}.

To enhance the security of NCSs, sophisticated cryptographic methods have been developed by {Information Technology (IT)} researchers. However, such techniques can introduce time delays and quantization errors \cite{nguyen2024observer}, which can potentially cause instability \cite{ali1998stability}. NCS also has strict performance requirements that are difficult to achieve under time delays \cite{ai2016stability}. Additionally, unlike IT security, which deals with data breaches, security in NCS must address both data breaches and potential physical consequences. In other words, an attack on an NCS can not only steal confidential information but also change the behavior of physical processes. To prevent such consequences, researchers and engineers do need ways to quantify how secure an NCS is, i.e., how easily the system can be attacked, how much damage can be caused, and how quickly it can recover. To this end, NCS security has received increased research interest from the control community \cite{chong2019tutorial}, which will be the focus of this review article.

\begin{figure}
    \centering
    \includegraphics[width=8cm]{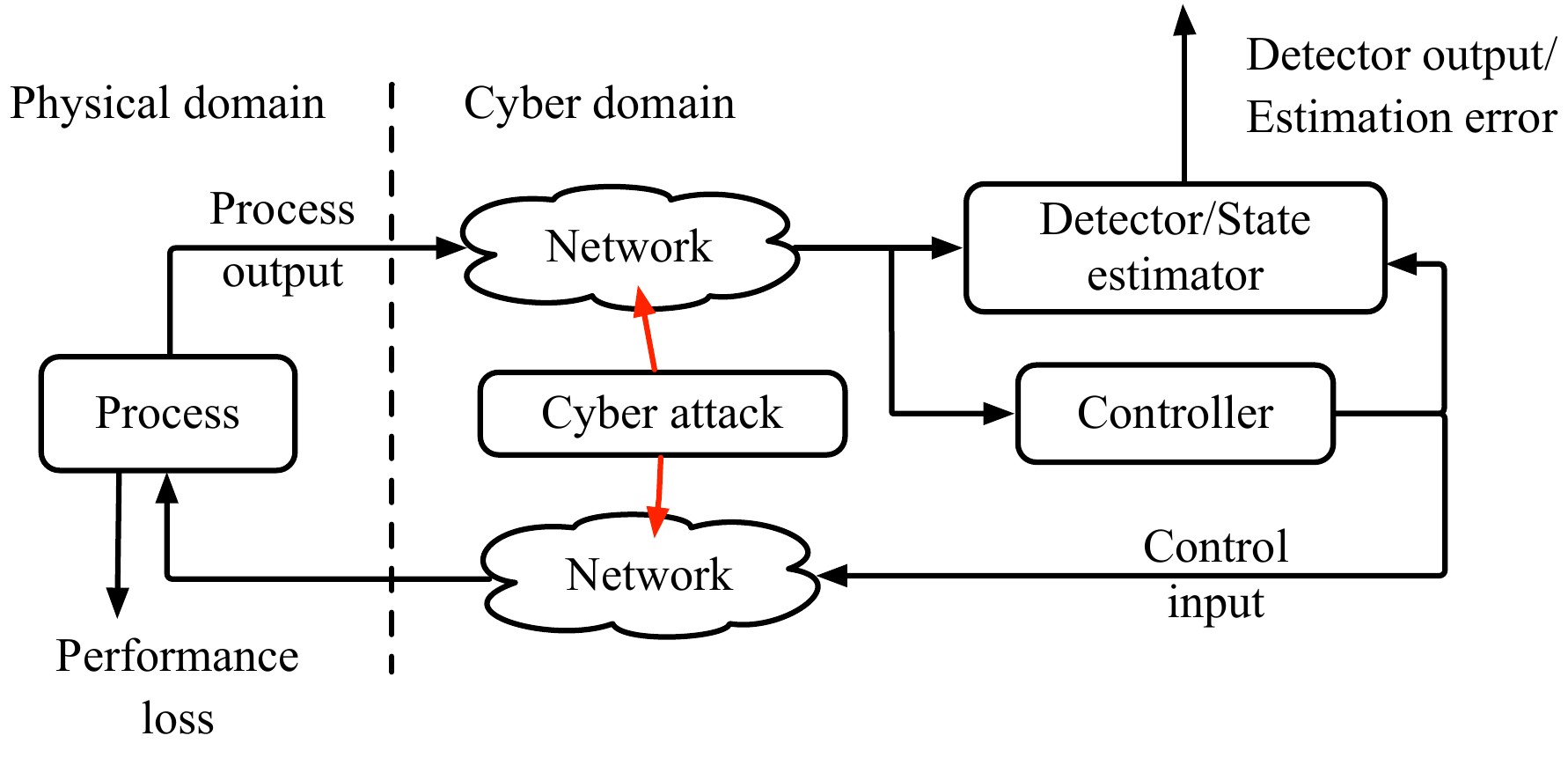} 
    \caption{{An} NCS under sensor and actuator attacks.}
    \label{fig:NCS}
\end{figure}

Typically, an NCS (see Figure~\ref{fig:NCS}) consists of a physical process on one side of the network and a controller and detector on the other side. The sensor data from the physical process are sent over a wireless communication channel to the controller and detector for control and monitoring. As mentioned before, the communication channels used in {NCSs} are generally (but not limited to) radio waves and the internet \cite{milovsevic2020security_thesis}, which might be accessible by adversaries. 

To mitigate such adversaries, the security of NCS has been reviewed from various perspectives in the past. An overview of the works can be found in Table~\ref{tab:review}, most of which adopt a control-theoretic viewpoint. In parallel, security has also been studied from neighboring communities such as computer security, systems engineering, and industrial automation. For example, \cite{cardenas2008research} provides an early overview of NCS vulnerabilities from a cybersecurity standpoint, while \cite{humayed2017cyber,blokland2020concepts} presents taxonomies of threats and defenses grounded in information security and system architecture. Industry-focused reviews such as \cite{knowles2015survey} emphasize operational risks and attack scenarios drawn from real-world incidents. However, these works often abstract away the dynamic, feedback-driven structure that defines NCS. Thus, in this review, we revisit NCS security from a control-theoretic viewpoint. \begin{quote}
In particular, we examine one particular aspect of NCS security: \emph{quantifying security}.
\end{quote}
\begin{table*}[ht]
\centering
\caption{Selected Survey and Review Works on NCS Security}
\label{tab:review}
\begin{tabular}{||p{3cm} | p{12cm}||} 
\hline
\textbf{Reference(s)} & \textbf{Description} \\
\hline
\cite{chong2019tutorial,how2015cyberphysical,giraldo2018survey,dibaji2019systems,sandberg2022secure} & Tutorial-style introductions to NCS security from a control-theoretic perspective. \\
\hline
\cite{sanchez2019bibliographical} & Reviews bibliographical definitions and taxonomies of cyber-physical attack types. \\
\hline
\cite{chen2021cyber,arauz2022cyber} & Discuss security concerns in distributed model predictive control frameworks. \\
\hline
\cite{huang2022reinforcement} & Explores security challenges in reinforcement learning-based NCSs. \\
\hline
\cite{ishii2022overview,liu2024survey} & Investigate vulnerabilities in consensus protocols within multi-agent systems. \\
\hline
\cite{brdys2014integrated} & Proposes mitigation strategies based on system reconfiguration in response to attacks. \\
\hline
\cite{li2023critical} & Examines the cyber-physical security of smart building control systems. \\
\hline
\cite{liu2022secure} & Applies formal methods to enable secure-by-design approaches in NCS. \\
\hline
\end{tabular}
\end{table*}

\begin{figure}
    \centering
    \includegraphics[width=7cm]{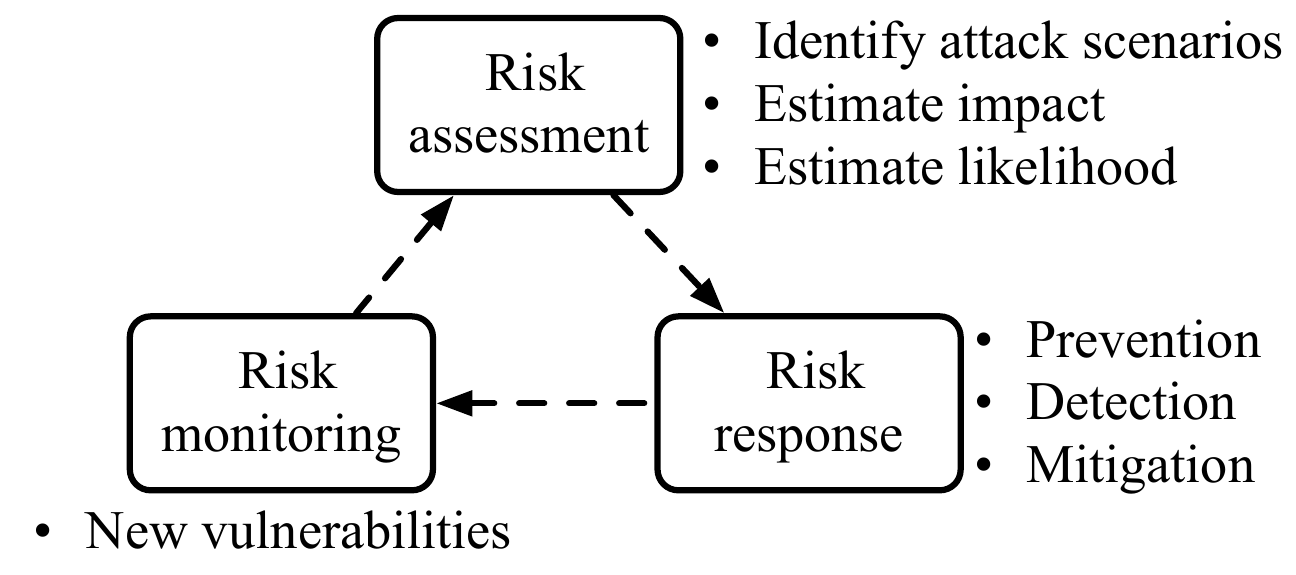} 
    \caption{Secure control system design approach using risk management framework adopted from \cite{ross2012guide,milovsevic2020security_thesis}. Identifying attack scenarios is discussed in \cite{teixeira2015secure}, methods to estimate impact are discussed in Section~\ref{sec:impact}, and methods to estimate attack likelihood are discussed in Section~\ref{sec:security}. Risk response strategies are discussed in detail in \cite[Section~V]{chong2019tutorial}, whereas we discuss some mitigation strategies in Section~\ref{sec:mitigation}.}
    \label{fig:Flow}
\end{figure}

The need to quantify security is explained using the risk management framework (see Figure~\ref{fig:Flow}), where a lower value of security is related to higher attack impact or higher likelihood associated with a given attack scenario. The consensus in the risk management literature is that to manage risks (or attacks), the defender\footnote{Here, the terms defender and operator are used interchangeably.} must follow the three steps: risk assessment, response, and monitoring. The risk assessment step aims to identify attack scenarios, estimate likelihood, and estimate the impact (physical degradation) caused. The risk response step focuses on designing risk prevention, detection, or mitigation strategies. The risk monitoring step aims to identify new vulnerabilities. Thus, quantifying the impact (physical degradation) and likelihood is an important step toward designing a secure NCS. 

As noted in \cite{teixeira2015secure}, the likelihood of a stealthy attack is influenced by the minimum number of components that needs to be compromised by the adversary (see Example~\ref{exmp:resource}). This minimal set of components is hereafter referred to as \emph{resources}. Drawing analogies from computer security \cite{ben2013using}, the required adversarial resources serve as a proxy for attack feasibility: the fewer the resources needed, the higher the likelihood of an attack, and vice versa. The explicit modeling of adversarial resources is also a central theme in security research \cite{do2019role,bauer2007low}. Motivated by the above arguments, we formulate the problem addressed in this review article.

\begin{quote}
What methods are available in the literature to quantify the attack impact and resources? How can the quantified metrics be used to mitigate attacks? What are the related open research challenges?
\end{quote}

No other work in the literature systematically reviews the methods to quantify attack impact and resources. Related work was conducted in \cite{zhu2021cybersecurity}, where the authors present an overview of security challenges in robotic systems, highlighting the importance of \emph{quantitative metrics} for security. Game theory is presented as an effective framework to quantify security, and the work discusses existing game theoretic algorithms in a tutorial fashion {where the game payoffs are fully known}. 
{Alongside this progress,}
the present article reviews methods to model {and compute} the game payoffs. The work \cite{pendleton2016survey} reviews metrics for quantifying security from an IT perspective without considering the physical degradation, which is also dealt with in the sequel.

\subsection{Outline} 
\begin{figure*}
\vspace{-100pt}
    \centering
    \includegraphics[scale=0.35]{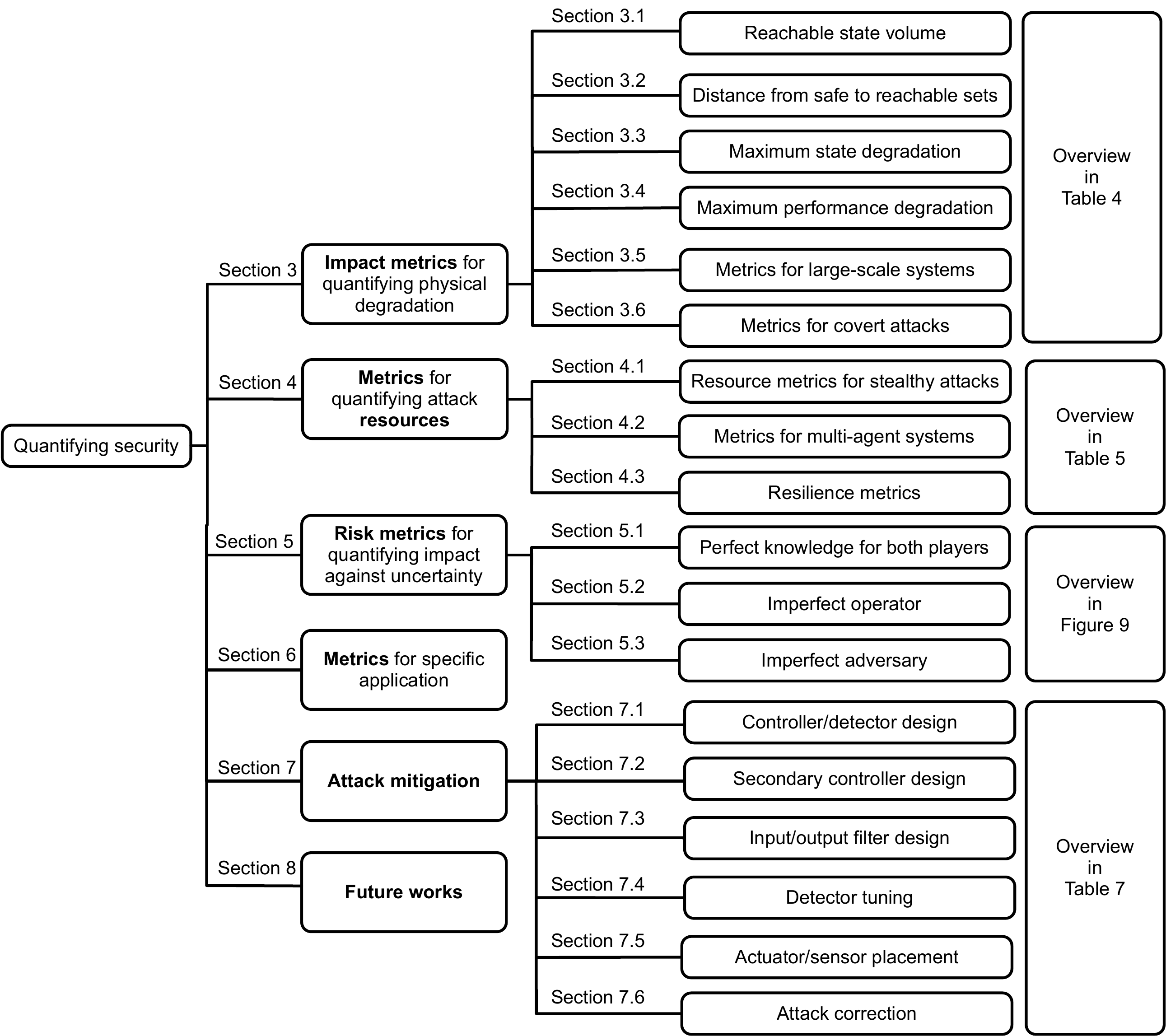}
    \caption{Graphical overview of this review article}
    \label{fig:overview}
    \rule{\textwidth}{0.4pt} 
    \vspace{-30pt}
\end{figure*}

The remainder of this article is organized as follows and is also depicted in Figure~\ref{fig:overview}. The system architecture and formal definitions of impact metric and resource metric are presented in Section~\ref{sec:problem}. In Section~\ref{sec:impact}, we review metrics that quantify the attack impact (physical degradation) caused by attacks. We review the metrics to quantify attack resources in Section~\ref{sec:security}. Probabilistic risk metrics are used to quantify the impact of adversaries with imperfect process knowledge in Section~\ref{sec:risk}. The impact and resource metrics developed for specific applications are discussed in Section~\ref{sec:application}. The attack mitigation strategies are detailed in Section~\ref{sec:mitigation}. We conclude the paper and provide avenues for future research in Section~\ref{sec:con}.
\subsection{Notation} 
Throughout this article, $\mathbb{R}, \mathbb{R}^{+}, \mathbb{C}, \mathbb{Z}$, and $\mathbb{Z}^{+}$ denote the sets of real numbers, non-negative real numbers, complex numbers, integers, and non-negative integers, respectively. The Kullback-Leibler (KL) divergence between the distributions $q_1$ and $q_2$ over a sample space $X$ is denoted as
\[
\mathcal{D}(q_1\|q_2) = \int\limits_X q_1(x)\log \frac{q_1(x)}{q_2(x)} \, \mathrm{d}x,
\]
where $\mathcal{D}(q_1\|q_2) = 0$ iff $q_1(x) = q_2(x), \forall x \in X$.

Given a vector $x \in \mathbb{R}^n$, let $x_i$ denote its $i^{\text{th}}$ element. We define the norms
\begin{align}
\|x\|_{\infty} &= \max_i |x_i|,\\
\|x\|_p &= \left( \sum\limits_{i=1}^{n} |x_i|^p \right)^{1/p}, \quad p > 0,\\
\|x\|_0 &= \#\{i \mid x_i \ne 0\},
\end{align}
and in words, $\|x\|_0$ represents the number of nonzero elements in the vector $x$. 

Let $x: \mathbb{Z} \to \mathbb{R}^n$ be a discrete-time signal, with $x[k]$ denoting its value at time step $k$. The $p$-norm of the signal $x$ over the time horizon $[0,N]$ is denoted by $\|x\|_{p,[0,N]}$. Similarly, the $\ell_p$-norm is denoted as
\[
\|x\|_{\ell_p, [0,N]} \triangleq \sum_{k=0}^{N} \|x[k]\|_p,
\]
and the $\ell_2$-norm over the horizon $[0,N]$ is denoted as
\[
\|x\|_{\ell_2, [0,N]}^2 \triangleq \sum_{k=0}^{N} x[k]^\top x[k].
\]
For simplicity, we denote the $\ell_2$-norm over the infinite horizon $[0,\infty]$ as $\|x\|_{\ell_2}^2$. The space of square-integrable signals is defined as
\[
\ell_2 \triangleq \left\{ x: \mathbb{Z}^+ \to \mathbb{R}^n \,\middle|\, \|x\|_{\ell_2}^2 = \|x\|^2_{\ell_2, [0,\infty]} < \infty \right\}.
\]
and the extended signal space be defined as 
\[
\ell_{2e} \triangleq \{ x: \mathbb{Z}^+ \to \mathbb{R}^n | \;||x||^2_{\ell_2,[0,N]} < \infty, \forall N \in \mathbb{Z}^+ \}\]

Given a square matrix $A \in \mathbb{R}^{n \times n}$, the largest singular value of $A$ is denoted by $\bar{\sigma}(A)$, and $\mathrm{trace}(A)$ denotes the sum of its diagonal elements.

Given a discrete-time transfer function $G(z)$, its norms are defined as
\begin{align}
\|G\|_2 &= \sqrt{\frac{1}{2\pi} \int_{-\pi}^{\pi} \mathrm{trace} \left[ G(e^{j\omega})^* G(e^{j\omega}) \right] \, d\omega},\\
\|G\|_\infty &= \sup_{\omega \in [0, 2\pi]} \bar{\sigma} \left( G(e^{j\omega}) \right),
\end{align}
where $(\cdot)^*$ denotes the conjugate transpose.
\begin{table*}
\vspace{5pt}
\caption{Overview of detection logics. Here $\Sigma_r$ represents the variance of the residues when there is no attack, $r_0$ represents the residual signal when there is no attack, CUSUM refers to \textbf{CU}mulative \textbf{SUM}, AMW refers to \textbf{A}verage over a \textbf{M}oving \textbf{W}indow, and MEWMA refers to \textbf{M}ultivariate \textbf{E}xponentially \textbf{W}eighted \textbf{M}oving \textbf{A}verage. The $\chi^2$ detector is a special case of the MEWMA detector when $\beta =1$. From samples $r_0[\cdot]$ and $r[\cdot]$, the KL divergence can be estimated efficiently \cite{wang2009divergence}.}
\label{tab:detectors}
\begin{center}
\begin{tabular}{||c | c | c||} 
 \hline
Name & Detection logic & Design parameters \\ [0.5ex] 
 \hline\hline
 CUSUM detector \cite{murguia2016characterization} & ${\begin{aligned}
 S^{+}[0]&=0,\; S^{-}[0]=0\\
 S^{+}[k+1] &= \max\left( 0,S[k]+ r[k]-b \right)\\
  S^{-}[k+1] &= \max\left( 0,S[k]- r[k]-b \right)\\
\min\left(S^{+}[k],S^{-}[k]\right) &\geq \tau \implies \text{alarm}\end{aligned}}$ & $\begin{aligned} \text{two}&\text{-sided CUSUM test}\\b &\text{- bias}\\ \tau &\text{- threshold}\end{aligned}$ \\
 \hline
\rule{0pt}{12pt} $\chi^2$ detector \cite{murguia2018reachable} &  $\displaystyle r[k]^T\Sigma_r^{-1}r[k]  \geq \tau \implies \text{alarm}$  & $\tau$ - threshold  \\
 \hline
   AMW p-detector \cite{akowuah2021real} & $ \sum_{t=k-L+1}^{k} \Vert r[t]\Vert_p \geq \tau \implies  \text{alarm}$  & $\begin{aligned} L &\text{- window length}\\ \tau &\text{- threshold}\end{aligned}$ \\
   \hline
   $\ell_p$ detector \cite{teixeira2015strategic} & $ \Vert r \Vert_{\ell_p} \geq \tau \implies  \text{alarm}$  & $\tau \text{- threshold}$\\
   \hline
     KL detector \cite{zaman2022security} & $ \mathcal{D}(r_0[k-L:k]\Vert r[k-L:k]) \geq \tau$ & $\begin{aligned} L &\text{- window length}\\ \tau &\text{- threshold}\\
     r_0 & \text{- nominal residues}\end{aligned}$  \\
   \hline
   \rule{0pt}{16pt}
 MEWMA detector \cite{umsonst2022experimental} & ${\begin{aligned} 
 x_D[0]&=0,\;x_D[k+1] = \beta \Sigma_r^{-0.5}r[k] + (1-\beta) x_D[k]\\
 y_D[k+1] &= \frac{2-\beta}{\beta} \Vert x_D[k+1]\Vert_2^2\\
 y_D[k] &\geq \tau \implies  \text{alarm}
  \end{aligned}}$  & $\begin{aligned} \beta &\in (0,1]\\ \tau &\text{- threshold}\end{aligned}$ \\[1ex]
   \hline
\end{tabular}
\end{center}
\end{table*}

%% file: Problem.tex
In this section, we describe the structure of the closed-loop NCS, the adversarial policy, and the detection scheme. We then formulate the problem studied. 
\subsection{System description}
We consider a linear time-invariant (LTI) discrete-time (DT) plant/process described as 
\begin{equation}\label{P}
\text{Process:} \left\{    \begin{aligned}
        x[k+1] &= Ax[k] + B \tilde{u}[k] + \omega[k]\\
                    y[k] &= C {x}[k] + v[k]\\
                    y_p[k] &= C_px[k] 
    \end{aligned}\right.
\end{equation}
where $x \in \mathbb{R}^{n}$ is the state of the plant,  $y \in \mathbb{R}^{m}$ is the measurement from the sensors, $\tilde{u} \in \mathbb{R}^{q}$ is the control input applied, and $y_p \in \mathbb{R}^{s}$ is the virtual performance loss output. Similar to robust control, the performance of the process is said to be good when the performance loss output is close to zero, e.g., tracking error. The i.i.d process noise and measurement noise are represented by $\omega \in \mathcal{N} (0,\Sigma_w)$ and $v \in \mathcal{N} (0,\Sigma_v)$, respectively. All the matrices are of appropriate dimension.
\begin{assumption}
The tuple $(A,B)$ is controllable and the tuple $(A,C)$ is observable. $\hfill \triangleleft$
\end{assumption}
We consider the process to be controlled using a full-order LTI output-feedback controller 
\begin{equation}\label{C}
  \text{Controller:}\left\{    \begin{aligned}
    z[k+1] &= A_cz[k] + B_c \tilde{y}[k] \\
    u[k] &= C_cz[k] +D_c \tilde{y}[k]  
    \end{aligned} \right.
\end{equation}
where $z \in \mathbb{R}^{n_c}$ is the state of the controller which can include the tracking/estimation error estimates, $\tilde{y} \in \mathbb{R}^{m}$ is the measurement received by the controller, and ${u} \in \mathbb{R}^{q}$ is the control input generated by the controller. It is worth noting that the control input applied $\tilde{u}$ and the measurement received $\tilde{y}$ might be different from the designed control input $u$ and the measurement output $y$ due to cyber-attacks (see Figure~\ref{fig:NCS}).

To detect attacks, the operator employs a residue generator as follows
\begin{equation}\label{D}
\text{Detector:} \left\{
    \begin{aligned}
        \hat{x}[k+1] &= A\hat{x}[k] + B u[k] +Ky_r[k]\\
                    y_r[k]&= \tilde{y}[k] - C\hat{x}[k] \\
                    r[k] &= Vy_r[k]
    \end{aligned}\right.
\end{equation}
where $r\in \mathbb{R}^{d}$ is the residue evaluated to detect attacks, $\hat{x} \in \mathbb{R}^n$ is the state estimate, $\hat{y} \in \mathbb{R}^m$ is the output estimate, and $K$ and $V$ are detector gains which are {design parameters.} Since the plant and the controller are linear, $r \sim \mathcal{N}(0,\Sigma_r)$ where $\Sigma_r \succ 0$. 

A summary of the detection logic used in the literature to identify attacks using the residue signal \( r \) is provided in Table~\ref{tab:detectors}. To express the detection logic compactly, we adopt the following notation
\begin{equation}\label{eq:det:logic}
\begin{aligned}
    f_D(r) \leq \tau & \implies \;\;\text{no alarm},\\
    f_D(r) > \tau & \implies \;\;\text{alarm}
\end{aligned}
\end{equation}
where the function \( f_D(r) \) represents one of the detection logics from Table~\ref{tab:detectors}, and \( \tau \) is the detection threshold. The threshold \( \tau \) is designed to trade off between attack impact and false alarm rate~\cite{umsonst2022finite}. An alarm indicates that the operator believes an attack is occurring.
\begin{remark}
The matrix $V$ is designed with an empty null space so that any changes in $y_r$ affect the residual signal. The variance of the residue $\Sigma_r$ can be computed using the variances $\Sigma_w$ and $\Sigma_v$ \cite{milovsevic2019estimating}. $\hfill \triangleleft$
\end{remark}
\subsection{Adversarial description}
This section introduces the adversary, its resources, and its strategy.
\subsubsection{System knowledge}
In much of the literature, the adversary is commonly assumed to have full knowledge of the system dynamics, including the plant, controller, detector, and detection scheme. While this may not reflect all practical scenarios, where the adversary's knowledge can range from none to partial (see Section~\ref{sec:risk}), this assumption provides a useful worst-case benchmark for evaluating detection performance. We adopt this assumption for consistency with existing works unless otherwise specified. Works that relax this assumption are discussed in Section~\ref{sec:risk}. 

\begin{assumption}\label{ass:knowledge}
The adversary is assumed to know the matrices in \eqref{P}, \eqref{C}, \eqref{D}, the detection function \eqref{eq:det:logic}, and the threshold \( \tau \). \hfill \( \triangleleft \)
\end{assumption}
\subsubsection{Disclosure and disruption resources}
Similar to computer security, we assume that the adversary has certain resources in order to conduct an attack. In particular, we consider that the adversary can access some of the sensor/actuator channels and read and alter the transmitted sensor/control data. In other words, we consider that 
\begin{equation}\label{eq:EF}
\begin{bmatrix}
\tilde{u}[k]\\
\tilde{y}[k]
\end{bmatrix} = 
\begin{bmatrix}
{u}[k]\\
{y}[k]
\end{bmatrix} +
\begin{bmatrix}
E_a & 0\\
0 & F_a
\end{bmatrix}
\begin{bmatrix}
    a_u[k]\\
    a_y[k]
\end{bmatrix}
\end{equation}
where $a_u \in \mathbb{R}^{q}$ and $a_y \in \mathbb{R}^{m}$ are the attack signals injected by the adversary into the actuator and sensor channels, respectively. Such attack strategies are called False Data Injection (FDI) attacks \cite{teixeira2015secure}. In this paper, we mainly focus on FDI attacks since it is widely studied in the literature. Some results pertaining to Denial-of-Service (DoS) attacks are discussed in Section~\ref{subsec:resilience}.

The matrix $E_a (F_a)$ in \eqref{eq:EF} is a diagonal matrix with $E_a(i,i)=1 \;(F_a(i,i)=1)$, if the actuator (sensor) channel $i$ is under attack and $0$ otherwise. A common assumption in the literature is that the adversary cannot simultaneously manipulate both sensor and actuator channels. While this assumption may not hold in all practical scenarios, it simplifies the analysis and reflects a more constrained adversary model. We adopt this assumption unless otherwise stated.

\begin{assumption}\label{ass:covert}
The adversary attacks either the sensors or the actuators, but not both. \hfill \( \triangleleft \)
\end{assumption}

If the adversary can access both the sensor and actuator channels, they may inject a covert attack~\cite{smith2015covert}. Under Assumption~\ref{ass:knowledge}, such attacks are provably undetectable. The impact of covert attacks under imperfect adversarial knowledge is explored in Section~\ref{subsec:covert:impact}.
\subsubsection{Attack constraints}
When there is no attack, it follows that $r \sim \mathcal{N}(0,\Sigma_r)$ where $r$ is the residue signal in \eqref{D}. However, under attack, the mean of the residue signal deviates from zero. If the residue signal deviates sufficiently, it triggers an alarm at the detector. Due to the fact that an adversary, who is detected, can be mitigated effectively, we consider a worst-case stealthy adversary that injects an attack signal {such that no alarm is raised.}
\begin{definition}[Stealthy attack/adversary]
An attack signal is defined as \emph{stealthy} if $f_D(r) \leq \tau$. An adversary that injects a stealthy attack is called a \emph{stealthy adversary}. \hfill $\triangleleft$
\end{definition}

In this paper, we consider stealthy adversaries that satisfy $f_D(r) \leq \tau$ where $\tau$ is the alarm threshold defined in \eqref{eq:det:logic}. Some works consider an adversary that satisfies $f_D(r) \leq \tau + \delta \tau$, where $\delta \tau$ is the adversary's willingness to risk detection. Interested readers are referred to \cite{milovsevivc2017analysis} for more details. Next, we show the importance of studying stealthy attacks through a numerical example.
\input{Example_tank}

\subsubsection{Attack policy}
\label{sec:attack_policy}
In general, two different adversaries are considered in the literature: (a) a maximum disruption adversary, which helps us to quantify attack impact, and (b) a minimum resource adversary, which is related to attack likelihood. We next formulate the attack policy for the aforementioned adversaries.

\emph{Maximum disruption stealthy adversary:} We consider an adversary that injects a stealthy attack signal to worsen the closed-loop system's performance. Let the performance of the closed-loop system be governed by $\mathcal{J}$, which is generally convex in the attack signal $a$. Here, $\mathcal{J}$ can represent the norm of the state-estimation error $e[k] = x[k] - \hat{x}[k]$, or tracking error $e[k] = y[k] - r^{*}$ over a given horizon length, where $r^*$ represents a given constant reference signal. Then, we consider an adversary that injects an attack signal by solving the optimization problem
\begin{equation}\label{eq:impact}
\boxed{
I \triangleq \left\{\begin{aligned}
\sup_{a} & \quad \mathcal{J}\\
\text{s.t.} & \quad f_D(r) \leq \tau\\
& \quad \eqref{P}-\eqref{eq:EF}
\end{aligned}\right.}
\end{equation}
where $a$ represents $a_u$ or $a_y$ depending on a sensor or actuator attack scenario considered, and $I$ represents the maximum performance loss caused.
If the value of the performance loss $I$ is high (low), then the NCS is non-robust (robust) against attacks. Henceforth, the optimization problem \eqref{eq:impact} is referred to as impact metric.
\begin{definition}[Impact metric]
The optimization problem in \eqref{eq:impact}, which maps the system dynamics in \eqref{P}-\eqref{eq:det:logic} to the performance loss caused by a stealthy adversary, is defined as an impact metric. $\hfill \triangleleft$
\end{definition}

\emph{Minimum resource stealthy adversary:} Let us consider that the adversary injects an attack $a_{y_j}$ into the sensor channel $y_j$. To remain stealthy, the adversary also has to inject attacks into some of the other sensor channels. The necessity to corrupt additional channels to remain stealthy might not be immediately apparent to the readers. To this end, we explain using an example.
\begin{exmp}
Consider a DC motor where the sensor measurements of speed (revolutions per minute) and torque (Newton meter) are transmitted over the network to a controller. In general, DC motors have a torque-speed operating curve. In other words, given a motor speed, the corresponding maximum torque should lie below a particular value. Then, if the adversary alters only the value of the speed by injecting an attack signal, such attacks can be easily detected. This is because the operator can infer that for the given speed measurements, the torque values do not satisfy the limits of the operating curve. Thus, in order to remain stealthy, the adversary also has to attack the torque values, demanding the adversary to corrupt additional channels. $\hfill \triangleleft$
\end{exmp}
Thus, given an adversary attacking a sensor channel $y_j$, other channels must be additionally corrupted such that the adversary remains stealthy. A minimum resource adversary maintains stealthiness by corrupting the least number of sensor channels. In other words, we consider an adversary that injects attack signals by solving the following optimization problem 
\begin{equation}\label{eq:security}
\boxed{
S_{j} \triangleq \left\{\begin{aligned}
\inf_{a_y} & \quad \Vert  a_y \Vert_p\\
\text{s.t.} & \quad f_D(r) \leq \tau \\
& \quad \Vert a_{y_j} \Vert_p \geq \epsilon, 1 \leq j \leq m\\
& \quad \eqref{P}-\eqref{eq:EF}
\end{aligned}\right.}
\end{equation}
where $S_{j}$ represents the least resources needed by the adversary to maintain stealthiness when sensor $y_j$ is under attack, and 
$\epsilon \in \mathbb{R}^{+}$ represents the {minimum} energy of the attack injected into sensor channel $y_j$. {If the value $S_{j}$ is high (low), then the attacker has to corrupt a high (low) number of sensor channels to stay stealthy. Thus, the value $S_j$ relates to the difficulty of maintaining stealthiness when sensor channel $j$ is under attack. In other words, the value of $S_{j}$ relates to the inverse of the attack likelihood of sensor channel $j$.} We next provide the definition of a resource metric after which we depict the relation between attack likelihood and attack resources using a numerical example.
\begin{definition}[Resource metric]
The optimization problem in \eqref{eq:security}, which maps the system dynamics \eqref{P}-\eqref{eq:det:logic} to the {minimum} resources needed by the stealthy adversary is defined as a resource metric. $\hfill \triangleleft$
\end{definition}
\begin{exmp}[\cite{milovsevic2020actuator}]\label{exmp:resource}
Consider the benchmark IEEE 14-bus power network depicted in Figure~\ref{fig:ieee14bus_rv}. The network is controlled using $N=5$ generators, and the operator has access to the measurements $\theta_1,\theta_3,\theta_5,\theta_7,\theta_9$ and $\theta_{11}$. Here, $\theta_i$ is the voltage angle of bus $i\in \{1,\dots,14\}$. We also consider the loads at bus $2,5,9$, and $14$ as actuators since they have considerable effect on the network. As mentioned in \cite{tegling2018fundamental}, the network can be described by an LTI system around a stable operating point. 

\begin{figure}[!b]
    \centering
    \includegraphics[width=\linewidth]{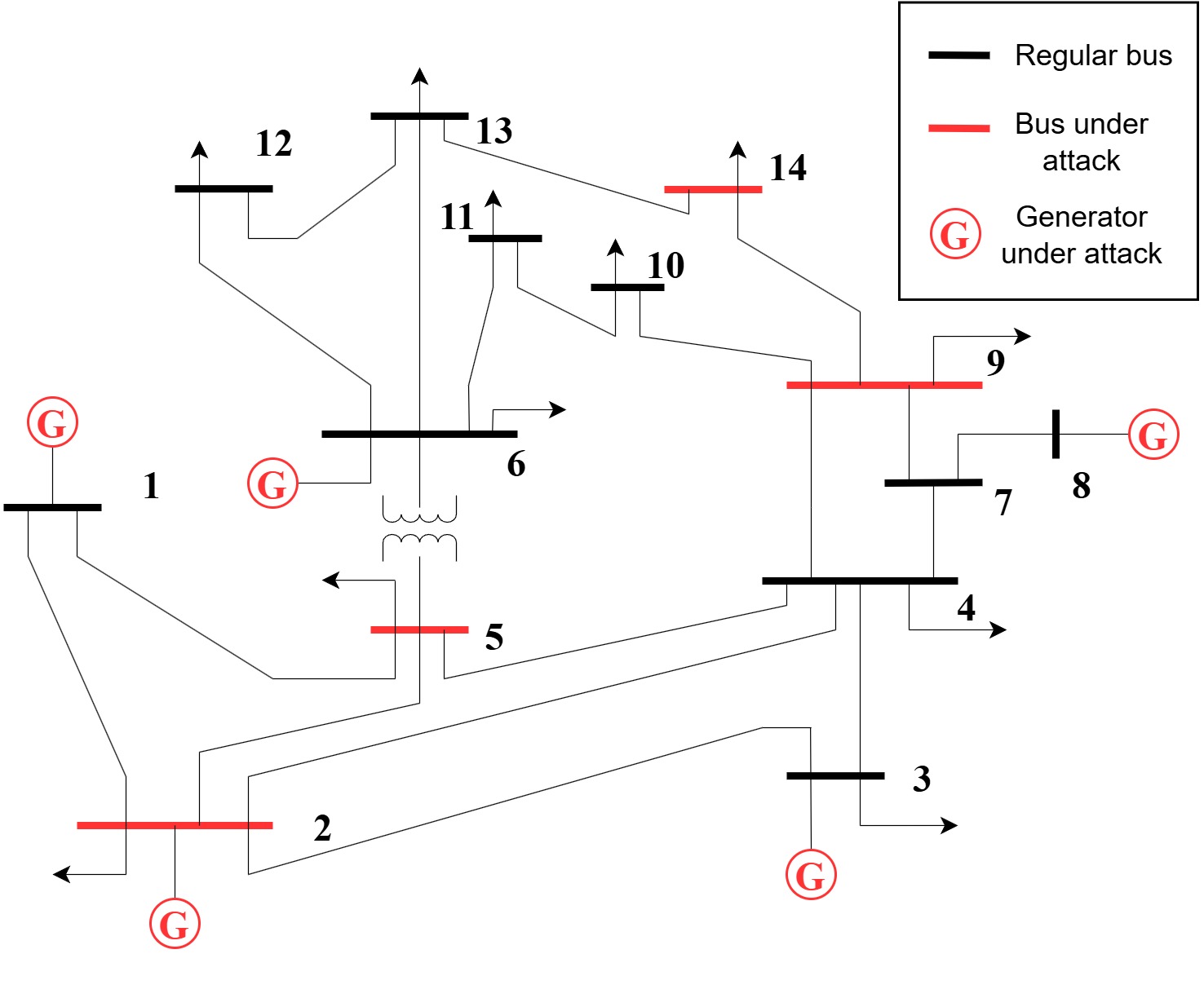}
    \caption{A pictorial representation of the IEEE 14-bus network under attack, where buses 2, 5, 9, and 14; and all the generators are attacked.}
    \label{fig:ieee14bus_rv}
\end{figure}

Now, given that a generator (actuator) $G_j, j \in \{1,\dots,5\}$ is under attack, we aim to determine the minimum amount of additional actuators that needs to be corrupted in order to conduct a perfectly undetectable attack, i.e., $r=0$. In other words, we aim to solve the optimization problem \eqref{eq:security} where the detection constraint is replaced with $r=0.$ The results are given in Table~\ref{tab:res:14}.
\begin{table}
\caption{Value of resource metric for IEEE 14-bus power network.}
\vspace{-10pt}
\label{tab:res:14}
    \centering
\begin{tabular}{ || p{1.5cm} | p{1.5cm}| p{1.5cm}|| }
\hline
Attacked generator & Resource metric $S_j$ & Attack likelihood\\\hline
$G_1$ & $4$ & Medium\\
$G_2$ & $2$ & High\\
$G_3$ & $5$ & Low\\
$G_4$ & $4$ & Medium\\
$G_5$ & $6$ & Low\\\hline
\hline
\end{tabular}
\vspace{-10pt}
\end{table}

As mentioned before, the resource metric relates to the \emph{effort} the adversary needs to invest to conduct a stealthy attack ($r=0$). Thus, for a minimum-resource stealthy adversary, the resources related to the inverse of the attack likelihood. The less the number of resources required, the higher the attack likelihood. $\hfill \triangleleft$
\end{exmp}
\subsection{Problem statement}
The objective of this paper is to provide a comprehensive overview of the methods from the literature to determine the value of the impact/resource metric in \eqref{eq:impact} or \eqref{eq:security}. Once these values are determined, the operator can improve security by minimizing {\eqref{eq:impact} or} maximizing {\eqref{eq:security}} through various techniques reviewed in Section~\ref{sec:mitigation}. 

%% file: Example_tank.tex
\begin{exmp}
\begin{figure}
    \centering
    \includegraphics[width=7cm]{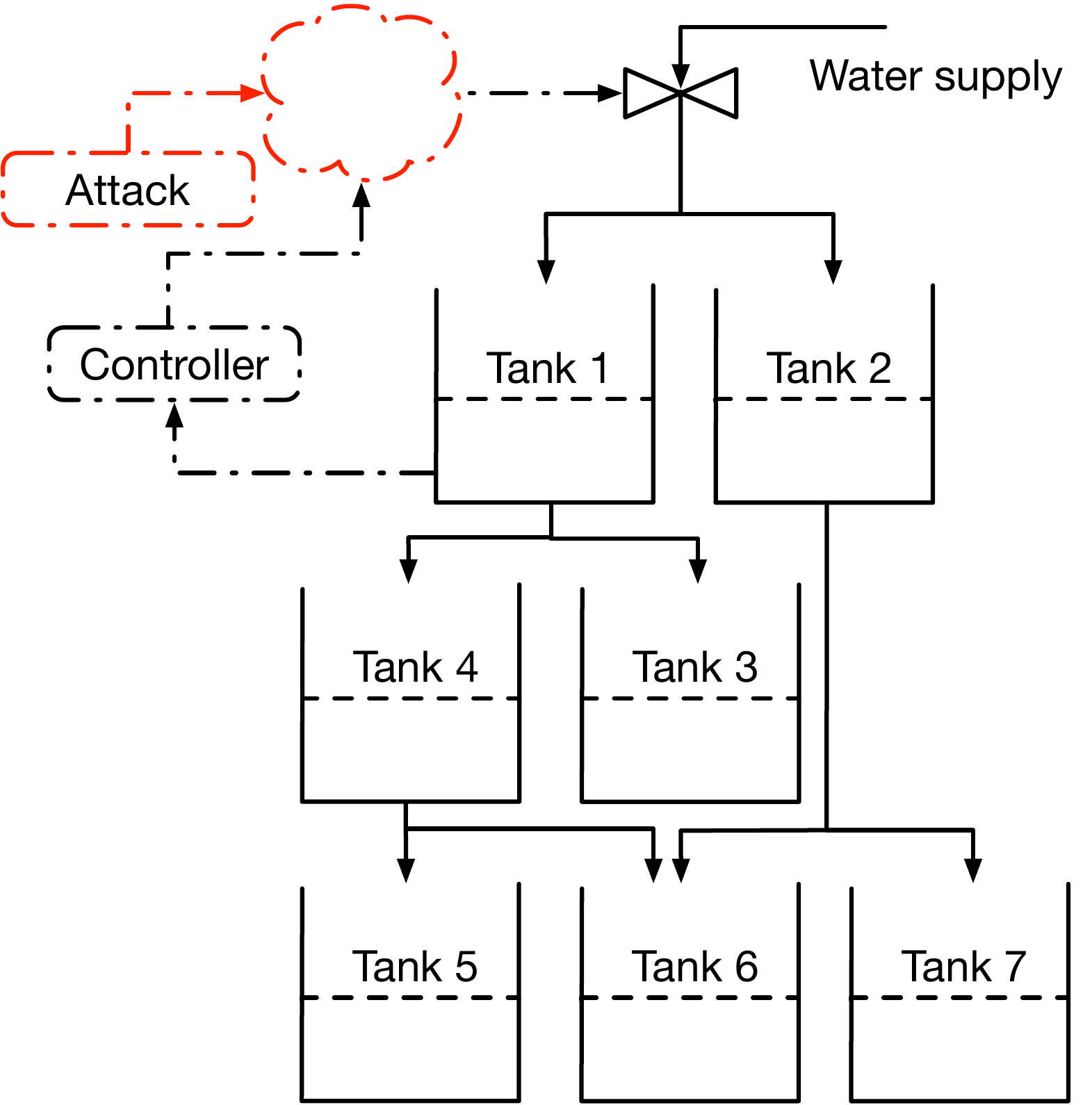} 
    \caption{Pictorial representation of interconnected reservoir system under an actuator attack. The solid lines represent the physical components/connections. The dashed-dotted lines represent the cyber components. The controller is designed such that \emph{Tank\;1} retains $r$ (in absolute units) amount of water.}
    \label{fig:7_tank}
\end{figure}
Let us consider an interconnected reservoir system as represented in Figure~\ref{fig:7_tank} \cite[Example 17]{farina2000positive}. Let ${x}_i, 1 \leq i \leq 7$ denote each tank's state, representing the amount of water (in absolute units) retained in the tank. For convenience, we denote the state of the entire system as $x \triangleq [x_1, x_2, \ldots, x_7]^T$.
Let $k_i, 1 \leq i \leq 7$ denote the parameter of each tank that represents the capacity of tank $i$ to retain water. The pipeline that distributes the water from the supply/tank to other water tanks is called a distribution line. There are $4$ distribution lines in Figure~\ref{fig:7_tank}, and $\alpha_j, 1 \leq i \leq 4$, $\alpha_j$ represents the ratio of water distributed between the output tanks. 

The operator can measure the water level in tank~$1$ and control the input supplied to tanks~$1$ (and tank~$2$) using a proportional controller. The performance of the reservoir system is evaluated based on the deviation of the water level in tank~$6$ from a reference level $\bar{r}$. Thus, the controller gain $K_P$ is designed to maintain a constant water level in tank~$1$, which in turn helps maintain the reference level $\bar{r}$ in tank~$6$.

We consider an adversary injects attacks into the actuator channel. Since the operator can measure the water level in tank $1$, an $\ell_2$ detector (see Table~\ref{tab:detectors}) is deployed in tank~$1$. In other words, an alarm is raised if $\vert x_1 \vert_{\ell_2} > \tau $. 

\begin{figure}
    \centering
    \includegraphics[width=8cm]{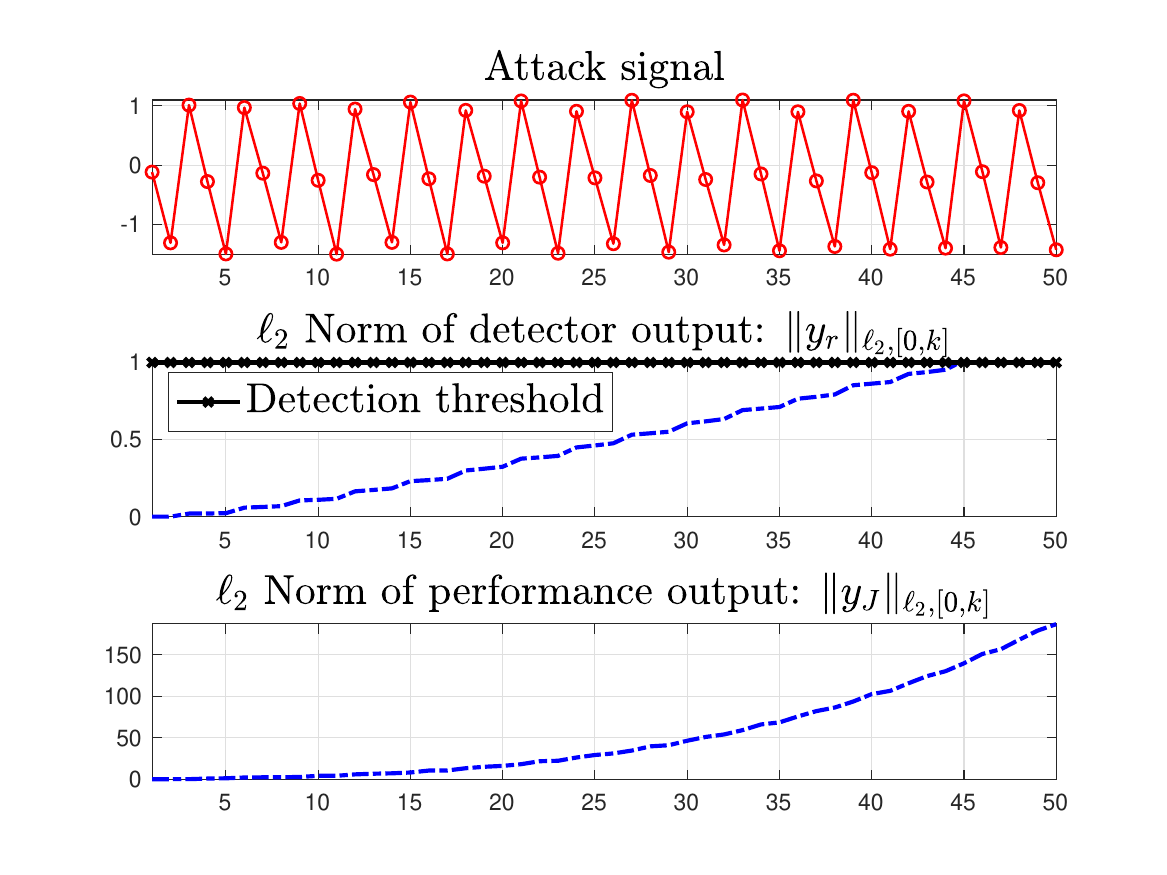}
    \vspace{-10pt}
    \caption{(Top) The attack signal in \eqref{eq:step} (Middle) $\ell_2$ norm of the detection output $r$, and the detection threshold $\tau$ (Bottom) $\ell_2$ norm of the performance output $y_J$.}
    \vspace{-15pt}
    \label{fig:step}
\end{figure}

To recall, we aim to discuss the effect of a stealthy attack signal on the closed-loop system. Due to the linearity, the response of the closed-loop system can be given as the sum of the response due to the reference signal $r$, the attack signals $a$, and the initial condition $x[0]$. Since we only aim to characterize the performance loss caused by attacks, we disregard the reference signal and the initial condition for now. Then, the closed-loop system from the attack input to the detection output $y_r$ and the performance output $y_p$ can be represented as 
\begin{align}
x[k+1] &= A_cx[k] + B_ca[k],\label{eq:closed:loop}\\
y_p[k]&= C_px[k],\\
y_r[k] &= C_rx[k]
\end{align}
\begin{table}
\caption{Parameter for interconnected reservoirs.}
\vspace{-10pt}
\label{tab:param}
    \centering
\begin{tabular}{ ||c|c|c|c|c|c|| }
\hline
 $K_P$ & $1.3$ & $k_1$ & $0.999$ & $k_2$ & $0.2$\\ \hline
 $k_3$ & $0.3$ & $k_4$ & $0.1$ & $k_5$ & $0.9$\\ \hline
 $k_6$ & $0.999$ & $k_7$ & $0.99$ & $\alpha_1$ & $0.1$\\ \hline
 $\alpha_2$ & $0.1$& $\alpha_3$ & $0.9$ & $\alpha_3$ & $0.3$\\ \hline\hline
\end{tabular}
\vspace{-10pt}
\end{table}
where 
\begin{align}
A_c(1,1) &= k_1-\alpha_1K_P, \, A_c(2,2) = k_2-(1-\alpha_1)K_P, \\
A_c(3,3) &= k_3, \, A_c(4,4) = k_4, \, A_c(5,5) = k_5, \\
A_c(6,6) &= k_6, \, A_c(7,7) = k_7, \\
A_c(3,1)&=(1-\alpha_2)(1-k_1),\;
A_c(4,1)=\alpha_2(1-k_1),\\
A_c(6,2)&= \alpha_3(1-k_2),\;
A_c(7,2)=(1-\alpha_3)(1-k_2),\\
A_c(5,4)&= \alpha_4(1-k_4),\;
A_c(6,4)=(1-\alpha_4)(1-k_4),\\
&\hspace{-1.1cm}
\text{and all other entries of}~ A_c ~\text{are zero}, \\
B_c&=\begin{bmatrix}
\alpha_1 & (1-\alpha_1) & 0_{1 \times 5}
\end{bmatrix}^T\\
C_p&=\begin{bmatrix}
0_{1 \times 5}& 1 & 0
\end{bmatrix}, C_r=\begin{bmatrix}
1 & 0_{1 \times 6}
\end{bmatrix}, 
\end{align}
and the parameters are given in Table~\ref{tab:param}. 

Next, let us consider an attack signal of the form
\begin{equation}\label{eq:step}
    a[k] = 1-1.2mod(k,3)+0.1sin[k]
\end{equation}
where $mod(\cdot,\cdot)$ represents the modulus operation. The effect of the attack \eqref{eq:step} on the detection and performance output is represented in Figure~\ref{fig:step}. From Figure~\ref{fig:step} (middle), the attack is stealthy since the $\ell_1$ norm of the detection output is below the detection threshold of $\tau =1$. For such a stealthy attack, we can see that the performance loss in Figure~\ref{fig:step} (bottom) is significant, depicting the significance of studying stealthy attacks. $\hfill \triangleleft$
\end{exmp}

%% file: Impact.tex
\subsection{Volume of the reachable states}\label{subsec:reachable}
In control theory, the reachable set at a given time refers to the collection of all states that a dynamical system can reach from a specified initial condition, under all admissible control inputs, within a finite time interval. It follows that the reachable set is altered in the presence of an adversary.

To assess this impact, any measure of the reachable set such as volume, diameter, or surface area can serve as a metric for evaluating system security and resilience. These measures quantify the \say{spread} or \say{reach} of the system under potential attack.

\begin{figure}
    \centering
    \includegraphics[width=6cm]{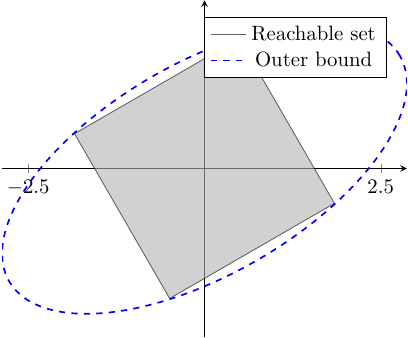}
    \caption{Pictorial representation of the reachable set of states and an ellipsoidal convex outer approximation discussed in Section~\ref{subsec:reachable}.}
    \vspace{-5pt}
    \label{fig:outer_approx}
\end{figure}

The work \cite{murguia2018reachable} considers an NCS with a $\chi^2$ detector and an adversary injecting stealthy attacks {into sensor measurements}. Let the set of all possible error trajectories ($e[k] \triangleq x[k] - \hat{x}[k]$) at time instant $k \in \mathbb{Z}$ be denoted as $\mathcal{R}$. Then, the main idea in \cite{murguia2018reachable} is to use the volume of the set $\mathcal{R}$ as a measure of impact (see Figure~\ref{fig:outer_approx}). 

The authors remark that determining the set $\mathcal{R}$ is a non-convex optimization problem in general. Thus, the authors derive an ellipsoidal outer approximation for the set $\mathcal{R}$, say $\mathcal{R}_e$. That is, the following
holds $\mathcal{R}_e\supseteq \mathcal{R}$ where 
\begin{equation}\label{eq:Re}
\mathcal{R} = \left\{ e[k] \in \mathbb{R}^{n} \Big| a_y \neq 0, (\ref{P})-(\ref{eq:det:logic}), \forall k \geq k^*
\right\}, 
\end{equation}
for any given $k^*$ and $\mathcal{R}_e$ is an ellipsoid. The authors then use the volume of the set $\mathcal{R}_e$ as a measure of impact. The extension of the work \cite{murguia2018reachable} proposes a method to derive the volume of the set $\mathcal{R}_e$ through a convex Semi-Definite Program (SDP) {(see \cite[Section 4.1]{murguia2020security})}. 

Then, for the metric proposed in \cite{murguia2020security} $\mathcal{J}$ and $f_D$ in \eqref{eq:impact} denotes the volume of $\mathcal{R}_e$, and the $\chi^2$ detector respectively. However, since the ellipsoidal approximation can be quite loose, the impact metric can be conservative. A geometric approach was proposed in \cite[Section III.B]{hashemi2018comparison} to overcomes the conservativeness. 

The concept of reachable states was first studied in \cite{mo2010false}, where the authors provided a method to compute the outer and inner approximations of the set of states reachable by the adversary under a KL detector. Set-based metrics were also adopted in \cite{vlahakis2023quantifying} to quantify the impact on switching systems, and in \cite{mo2015performance} to study the impact of joint sensor and actuator attacks. A volumetric metric was also adopted in the IT research field \cite{iganibo2023attack}.

Reachable sets can also be computed for nonlinear systems.
The work \cite{lin2023secondarysos} proposes a Sum-of-Squares (SOS) program to determine the forward reachable set for polynomial systems. The authors extend their work to sector-bounded non-linearities in \cite{lin2025secondary}.
\subsection{Distance between the reachable and safe sets}
\begin{figure}
    \centering
    \includegraphics[width=6cm]{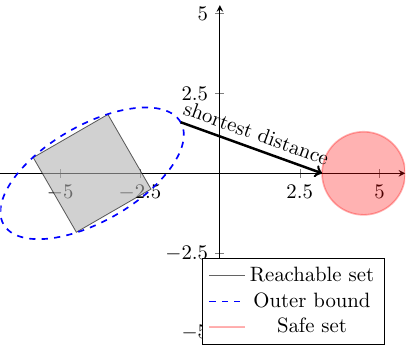}
    \caption{Pictorial representation of the shortest distance between the outer approximation of the reachable set and the safe set. }
    \vspace{-5pt}
    \label{fig:outer_dist}
\end{figure}

When {the collection of} the critical states of an NCS is defined as a safe set $\mathcal{S}_x$, the impact metric {is defined by} the minimum distance between the reachable set of states by the adversary ($\mathcal{R}_x$) and the set of critical states ($\mathcal{S}_x$) {\cite[Section 4.2]{murguia2020security}}. In other words, let us consider the NCS under sensor {attacks} and define the sets
\begin{align}
\mathcal{R} \supseteq \mathcal{R}_x &= \left\{ x \in \mathbb{R}^{n} \Big| a_y \neq 0, (\ref{P})-(\ref{eq:det:logic})\label{eq:reachset}
\right\},\\
\mathcal{S}_x &= \left\{ x \in \mathbb{R}^{n} \Big| x^TQx \leq 1 \label{eq:safeset}
\right\}.
\end{align}
Then for the metric proposed in \cite[Section 4.2]{murguia2020security}, $\mathcal{J}$ and $f_D$ in \eqref{eq:impact} denotes the inverse of the minimum distance between $\mathcal{R}_x$ and $\mathcal{S}_x$ and the $\chi^2$ detector respectively. As mentioned before, since the set $\mathcal{R}_x$ is non-convex, the authors use the distance between the ellipsoidal outer approximation of the set $\mathcal{R}_x$ and the $\mathcal{S}_x$ {as the impact} (see Figure~\ref{fig:outer_dist}). A similar metric was adopted in \cite{morris2017design}.

A binary security metric was proposed in \cite[Problem 2]{escudero2019prevention}. Here, the NCS is defined to be secure if the reachable set $\mathcal{R}_x$ does not intersect with a predefined dangerous set $\mathcal{D}_x$. For instance, the dangerous set can be defined as $\mathcal{D}_x = \mathbb{R}^n \backslash \mathcal{S}_x$, where $\mathcal{S}_x$ is defined in \eqref{eq:safeset}. 
\subsection{Maximum state degradation}\label{sec:error:degrade}
In \cite{murguia2016characterization}, the authors designate $m$ CUSUM detectors at the controller side, where $m$ is the number of sensor measurements. Then, \cite[Proposition 1]{murguia2016characterization} gives an upper bound on the Euclidean norm of the worst-case state deviation caused by a stealthy adversary. In other words, the work \cite{murguia2016characterization} determines the value of 
\begin{equation}\label{eq:states:diverge}
\lim_{k \to \infty} \big\vert \mathbb{E} [x[k]] \big\vert.
\end{equation}
Thus, for such a metric, $\mathcal{J}$ and $f_D$ in \eqref{eq:impact} denotes \eqref{eq:states:diverge} and a CUSUM detector, respectively. The work was also extended to a vector case, where all the sensor outputs are considered in a single detector \cite{murguia2016cusum,murguia2019model}. 

Instead of considering the maximum state degradation, another approach is to consider the worst-case error induced by any stealthy adversary. For instance, the work \cite{sui2020vulnerability} considers a metric similar to \eqref{eq:states:diverge} where 
\begin{equation}\label{eq:errors:diverge}
\mathcal{J} \triangleq \lim_{k \to \infty} \big\vert \mathbb{E} [e[k]] \big\vert.
\end{equation}
and $e[k]=x[k]-\hat{x}[k]$. The authors give a (possibly loose) upper bound of the maximum state degradation caused \cite[Theorem 3]{sui2020vulnerability}. The conditions under which the NCS is vulnerable to undetectable attacks are characterized in \cite[Theorem 1]{sui2020vulnerability}. Similar approaches were adopted in \cite{shinohara2018reach,wang2023worst}.

The work \cite{chen2017optimal} considers a setup where the adversary aims to drive the plant to a desired state $x^* \in \mathbb{R}^n$ in {a} finite horizon. Then, the authors quantify the error between the true state and the desired state $x^*$. Here, the adversary is not constrained to be stealthy; however, the innovations of the Kalman filter under attack are constrained to be instantaneously bounded.

Instead of considering a specific structure of the detection logic, the work \cite{bai2014kalman} considers a generic detection logic that conducts ergodicity-based tests on the residue signals (see \cite[Assumption 2]{bai2014kalman} for the definition of ergodicity-based tests). Then, the work provides closed-form expressions for the mean squared error (MSE) of the state estimate \cite[Theorem 1]{bai2014kalman} for a single-input single-output (SISO) scalar system. The authors {relax} the ergodicity assumption in \cite{bai2015security} and {provide} bounds on the worst-case MSE. 

A bound on the asymptotic estimation error {\eqref{eq:errors:diverge}} was derived for nonlinear systems in \cite[Theorem 1]{sargolzaei2021secure}. Here, the authors do not consider any detector but assume that the attack energy is bounded. Similarly, the work \cite{khazraei2022resiliency} considers a KL detector and derives the maximum state deviation an attacker can induce on a nonlinear system. 
\subsection{Maximum performance degradation}
The performance of the NCS 
is usually governed by the norm of a virtual performance loss signal ($y_p$ in \eqref{P}), similar to robust or Linear Quadratic (LQ) control. The performance of the closed-loop NCS is deemed to be good when the value of the performance loss signal is small (over a horizon). In this section, we consider an adversary that aims to maximize the performance loss. In other words, this section aims to quantify the maximum degradation of the virtual performance loss output in the presence of stealthy attacks. When the performance loss signal is defined as the state measurement error, the results in this section reduce to the results discussed in Section~\ref{sec:error:degrade}.  

This subsection is divided into three parts. In Section~\ref{sec:impact:finite}, we consider impact metrics that quantify security as the maximum performance degradation caused by stealthy attacks in the finite horizon. In Section~\ref{sec:impact:infinite}, we consider metrics that quantify the performance degradation caused in the infinite horizon, and in Section~\ref{subsec:stochastic}, we consider impact metrics with stochastic constraints or objective function.
\subsubsection{Impact metrics: Finite horizon}\label{sec:impact:finite}
The work \cite{teixeira2013quantifying} considers an $\ell_p$ detector in the finite horizon and an adversary injecting stealthy attacks into the sensors or actuators. The authors quantify impact as the value of the optimization problem \eqref{eq:impact} with 
\begin{equation}\label{eq:2012:Andre}
\mathcal{J} \triangleq \Vert y_p[k_0:k_0+N]\Vert_p, \; f_D(r) \triangleq \Vert r[k_0:k_0+N]\Vert_q.
\end{equation}

Then, \cite[Theorem~1]{teixeira2013quantifying} establishes the conditions under which the impact metric remains bounded (see Remark~\ref{rem:bounded}). For the case where $p = q = 2$, \cite[Theorem~2]{teixeira2013quantifying} provides a closed-form expression for the impact value, along with the corresponding attack signal. Related work along similar lines can be found in \cite{kwon2013security,wang2020optimal}.

The work \cite{milovsevic2017exploiting} quantifies the impact as the value of the optimization problem 
\begin{equation}\label{eq:Jez:inf}
\begin{aligned}
\sup_{a} & \quad \mathcal{J} \triangleq \Vert y_p[k_0:k_0+N]\Vert_\infty\\
\text{subject to} & \quad \underline{u} \leq {u}_a \leq \bar{u}, a_u=0
\end{aligned}
\end{equation}
where $u_a$ is the input generated by the feedback controller when the sensors are under attack, $\underline{u}$ and $\bar{u}$ are the control limits the attacker has to obey to maintain stealth. {It is worth noting that the stealthiness constraint in \eqref{eq:Jez:inf} can be cast similar to \eqref{eq:impact} when $f_D \triangleq |u_a - (\bar u + \underline{u})/2|$ and $\tau \triangleq (\bar u - \underline{u})/2$.} Here, the performance degradation is considered over a finite horizon of length $N$, and there are no explicit stealthiness constraint. The authors also show that the value of the problem \eqref{eq:Jez:inf} can be determined using a Linear Program (LP). 

The work \cite{milovsevic2018quantifying} considers an impact metric similar to \eqref{eq:Jez:inf} under three detectors: an AMW detector, a CUSUM detector, and a MEWMA detector (see Table~\ref{tab:detectors}). The authors provide an LP to determine the attack impact under the detectors mentioned above for many classes of attacks, such as DoS attacks \cite{amin2009safe}, routing attacks \cite{ferrari2017detection}, and replay attacks \cite{mo2009secure}. 

A metric similar to \eqref{eq:Jez:inf}, but based on an $\ell_{\infty}$ detector, was considered in \cite{hirzallah2018computation}. An $\ell_{\infty}$ detector constrains the maximum value of the residue signal under attack to remain below the detection threshold $\tau$. The authors formulate a linear program (LP) to compute the optimal attack vector and the corresponding impact.

The work \cite{shames2017security} quantifies impact as the weighted sum of the finite horizon $H_2$ norm from the attack input to the performance output and the $H_2$ norm from the attack input to the measurement output. The $H_2$ norm quantifies the norm of the process output when the input is a white noise. Thus, in \cite{shames2017security}, the authors consider the {adversary} injecting white noise and aim to measure the impact caused. Here, the authors do not consider any explicit detector. The value of the metric can be determined by solving an edge weight biclique problem, whose solution can be
approximated by a variety of methods from the literature \cite{dawande2001bipartite}.
\begin{remark}\label{rem:bounded}
It is critical to understand the different implications of an \emph{unbounded} value of the impact metric. Firstly, it indicates the existence of an attack vector capable of causing infinite performance degradation without detection. Thus, it exposes a critical vulnerability of the NCS. Secondly, it is not necessary that an unbounded impact can only be caused by an attack vector of unbounded energy. For instance, when the process is unstable, the attacker can cause infinite estimation error (see the discussion around equation (9)-(10) in \cite{umsonst2018game}). Finally, since the metric is unbounded, the corresponding design problem, which aims to design the controller gains to minimize attack impact, becomes ill-posed. Thus, the metric has to be amended to make the design problem well-posed. For instance, the works \cite{anand2020joint,teixeira2019optimal} propose an amended design problem for the ill-posed design problem in \cite{teixeira2015strategic}. $\hfill \triangleleft$
\end{remark}
\subsubsection{Impact metrics: Infinite horizon}\label{sec:impact:infinite}
Solving for impact metrics over a finite horizon has two main disadvantages. First, it requires computing Toeplitz matrices to evaluate the impact \cite[(12)]{teixeira2013quantifying}. As the horizon length increases, these Toeplitz matrices can become numerically ill-conditioned, particularly when the system dynamics are nearly unstable, resulting in poor numerical accuracy. Second, there are no general guidelines for selecting an appropriate horizon length for evaluating security.

To address these issues, this section reviews impact metrics that quantify performance degradation over an infinite horizon. However, the norm of the performance signal over an infinite horizon is unbounded in the presence of noise. Consequently, the works discussed in this subsection assume the absence of process and measurement noise (see Remark~\ref{rem:noise}).

The work \cite{teixeira2015strategic} extends the impact metric in \eqref{eq:2012:Andre}, which is defined over a finite horizon, to an infinite horizon setting. In other words, \cite{teixeira2015strategic} defines impact as the optimization problem \eqref{eq:impact} with 
\begin{equation}\label{eq:OOG}
\mathcal{J} \triangleq \Vert y_p\Vert_{\ell_2}^2, \; f_D(r) \triangleq \Vert r \Vert_{\ell_2}^2,
\end{equation}
whose value is referred to as the Output-to-Output Gain (OOG). Using dissipative system theory \cite{moylan2014dissipative} and strong duality \cite[Chapter 4]{petersen2012robust}, the value of the impact metric with \eqref{eq:OOG} is shown to be equivalent to an SDP \cite{anand2021risk}. 

The necessary and sufficient conditions for the value of the OOG to be bounded are defined in terms of unstable zeros of the NCS. More specifically, the value of the OOG is unbounded for a system with unstable zeros \cite[Theorem 2]{teixeira2015strategic}. Thus, if one is interested in re-designing the controller gain to minimize the attack impact, which is characterized by the OOG, the design problem would be infeasible for a closed-loop system with unstable zeros. In the following, we review two extended versions of the OOG, addressing this limitation. 

Firstly, an impact metric called the cyclic OOG was proposed in \cite{anand2023risktac} where the authors consider the optimization problem \eqref{eq:impact} with \eqref{eq:OOG} and the 
constraint 
\begin{equation}\label{eq:COOG}
\lim_{k \to \infty}\bar{x}[k]=0.
\end{equation}
where $\bar{x}[k]$
denotes the states of the closed-loop system. If the adversary injects an attack for a finite time horizon (say $N$), then the constraint \eqref{eq:COOG} is satisfied for any {asymptotically} stable closed-loop system (albeit $N$ being unknown). Thus, when the constraint \eqref{eq:COOG} is imposed on \eqref{eq:OOG}, it was shown in \cite{anand2023risktac} that the value of the optimization problem can be determined using an SDP; however, the boundedness conditions depend only on the zeros on the unit circle, not the unstable zeros. 

Secondly, a metric called the Attack energy-constrained OOG (AEC-OOG) was proposed in \cite{anand2023risk} where the constraint $\Vert a \Vert_{\ell_2}^2\leq \epsilon_r$ was added to the optimization problem \eqref{eq:OOG}. 
The value of the AEC-OOG can be determined using an SDP, and the metric is bounded if the closed-loop NCS is stable. 

The AEC-OOG was further extended to use the induced $\ell_1$ norm instead of the $\ell_2$ norm in \cite{anand2024scalable}. In other words, the authors consider an adversary that aims to maximize the absolute sum of the performance loss output when maintaining the absolute sum of the residue below a threshold. {More specifically, \cite{anand2024scalable} defines the impact as the optimization problem \eqref{eq:impact} with
\begin{align}
    \mathcal{J} \triangleq  \Vert y_p\Vert_{\ell_1}, \; f_D(r) \triangleq \Vert r \Vert_{\ell_1}.
\end{align}
}
When the closed-loop system is positive \cite{farina2011positive}, the AEC-OOG was proven to be equivalent to an LP. This is one of the few works that quantifies impact in the infinite horizon using an LP. Since the impact is determined using an LP instead of an SDP, the metric is also scalable. Similar scalability arguments were also made in \cite{rantzer2018tutorial}. 
\begin{remark}\label{rem:noise}
Let us consider the attack vector to be independent of the noise signals $\omega[\cdot]$ and $v[\cdot]$. Then, due to the superposition principle of linear system, we can denote $\bar{x}[k] = x_a[k]+ x_n[k]$ where $\bar{x}[k]$ is the state of the closed-loop NCS, and ${x}_a$ and $x_n$ denote the influence of attack and noise on the state, respectively. Thus, we can describe the closed-loop NCS as the summation of two systems. Once we compute the maximum performance degradation caused by attacks (say $\mathbb{I}_1$), we can compute the degradation caused by the worst-case process noise (say $\mathbb{I}_2$). Thus, the total impact is $\mathbb{I}_1+\mathbb{I}_2$. However, since $\mathbb{I}_2$ is not a function of $a[k]$, we assume that $\omega[k]=0, v[k] = 0 \;\forall \;k \geq 0$. $\hfill \triangleleft$
\end{remark}
\begin{remark}
Using strong duality, the OOG in \eqref{eq:OOG} can be rewritten as 
\vspace{-10pt}
\begin{equation}\label{eq:OOG:ref}
\sup_a \frac{ \Vert y_p \Vert_{\ell_2}^2}{\Vert y_r\Vert_{\ell_2}^2}. 
\end{equation}
The OOG in \eqref{eq:OOG:ref} can be easily misconstrued as the ratio of $H_{\infty}$ norm of two transfer functions such as 
\vspace{-5pt}
\begin{equation}\label{eq:OOG:ref2}
\sup_{\omega} \frac{\bar{\sigma}(G_{a \to y_p}(e^{j\omega}))}{\bar{\sigma}(G_{a \to y_r}(e^{j\omega}))}. 
\end{equation}
However, the value of \eqref{eq:OOG:ref} and \eqref{eq:OOG:ref2} are \emph{not equal}. Instead, the OOG is the ratio of the $H_{\infty}$ norm and the $H_{\_}$ index \cite{liu2005lmi}. More details about the relationship between \eqref{eq:OOG:ref} and \eqref{eq:OOG:ref2} can be found in \cite[Section 4.1.2]{teixeira2021security}. $\hfill \triangleleft$
\end{remark}
\subsubsection{Impact metrics: stochastic constraints or objective functions}\label{subsec:stochastic}
In this section, we consider impact metrics with probabilistic constraints or objective functions. For instance, \cite{milovsevivc2017analysis} considers an adversary that aims to increase the performance loss by injecting a constant bias into sensor channels. The adversary also aims to maintain the false alarm rate of the $\chi^2$ detector below a predefined threshold $\alpha$. Thus \cite{milovsevivc2017analysis} quantifies impact as the value of the optimization problem \eqref{eq:impact} where 
\begin{equation}\label{eq:Jez:error}
\mathcal{J} \;{\triangleq}\; \lim_{k \to \infty} |\mathbb{E}[e[k]]|, \;f_d(r)\; {\triangleq} \;\text{FAR of}\;\chi^2\;\text{detector}.
\end{equation}
The authors show that under mild assumptions, the value of \eqref{eq:Jez:error} can be determined using a quadratically constrained quadratic program (QCQP), for which efficient numerical solvers exist \cite{boyd2004convex}. A similar problem setup was also considered in \cite{gualandi2023worst}.

The work \cite{milovsevic2019estimating} quantifies impact as the mean performance degradation caused under a KL-detector, which can be formulated as the optimization problem \eqref{eq:impact} with
\begin{equation}\label{eq:Jez:inf:jnl}
\begin{aligned}
\mathcal{J} &\triangleq \mathbb{E}[\Vert y_p[k_0:k_0+N]\Vert_{\infty}],\\
 f_D(r) &\triangleq \mathcal{D}({r_0[k_0:k_0+N] \Vert r[k_0:k_0+N]}),
\end{aligned}
\end{equation} 
where $r_0$ represents the residual signal in the absence of attacks, and is determined apriori. The authors show that the optimal value of \eqref{eq:Jez:inf:jnl} can be obtained through a series of convex optimization problems \cite[Theorem 1, Algorithm 1]{milovsevic2019estimating}. When the objective function $\mathcal{J}$ in \eqref{eq:Jez:inf:jnl} is replaced with the asymptotic state estimation error, the corresponding impact is studied in \cite{fang2019stealthy}.

The paper \cite{milovsevic2019estimating} also considers a setup where impact is defined as the probability the states leave a safe set $\mathcal{S}_x$. In other words, \cite{milovsevic2019estimating} defines impact metric as the optimization problem \eqref{eq:impact} with 
\begin{equation}\label{eq:Jez:inf:jnl2}
\begin{aligned}
\mathcal{J} &\triangleq  \mathbb{P}[\vert y_p^i[k_0:k_0+N]\vert >1], \\
f_D(r) &\triangleq \mathcal{D}({{r}_0[{k_0:k_0+N}] \Vert r[{k_0:k_0+N}]}),\\
\end{aligned}
\end{equation}
where $y_p \in \mathbb{R}^q$. Although an algorithm to determine the exact value of \eqref{eq:Jez:inf:jnl2} is missing, \cite[Theorem 2]{milovsevic2019estimating} proposes a convex optimization problem to derive tight upper and lower bounds for the impact metric \eqref{eq:Jez:inf:jnl2}. The impact metric is also shown to be applicable to other classes of attacks, such as DoS and replay attacks. 

A metric similar to \eqref{eq:Jez:inf:jnl} was proposed for nonlinear time-delayed systems in \cite{wang2017security}. Here, the authors consider sector nonlinearities and the attack to be stochastic and bounded. In particular, the impact metric is defined as the optimization problem \eqref{eq:impact} where 
\begin{equation}\label{eq:non:Jez:E}
\mathcal{J} \triangleq \mathbb{E}[\Vert e[k]\Vert_{\infty}],
\end{equation} 
and there are no detection constraints. The authors provide state estimation error bounds using the attack energy, and noise bounds. 

The work \cite{sasahara2022attack} considers a nonlinear control system under actuator attacks. The authors model the dynamics using a finite horizon discrete Markov Decision Process (MDP) where the attacker can arbitrarily change the actions while remaining stealthy. Here, the adversary is defined as stealthy if the FAR does not exceed a predefined threshold. The authors assume that the states that trigger a detection alarm are predefined and known. The alarm states are embedded into the MDP, and the impact metric is defined as the optimization problem \eqref{eq:impact} where 
\begin{equation}
\mathcal{J} \triangleq \mathbb{E}\left[\sum_{k=0}^{T} R(k)\right],
\end{equation} 
and $R(k)$ is the reward received by the adversary, and the expectation is taken over all the admissible adversarial policies. The authors observe that the impact becomes computationally intractable. This is because the optimal attack policy depends not only on the current state but also on the entire history of states.
By augmenting the MDP states with the alarm state, \cite{sasahara2022attack} proposes a method to address the intractability. The work was extended to continuous MDPs in \cite{sasahara2023attack}.

The authors in \cite{ding2016security} consider a nonlinear NCS (nonlinear control affine process and controller) under bounded sensor and actuator attacks with known bounds and no stealthiness constraints. The paper then provides (nonlinear) sufficient conditions in \cite[Theorem 1]{ding2016security} for the closed-loop NCS to satisfy the safety constraint 
\begin{equation}\label{eq:non:Jez:P}
\mathbb{P}[\vert y_p^i[k_0:k_0+N]\vert <1].
\end{equation}
This result provides a probabilistic guarantee on safety under bounded nonlinear attacks.

In summary, the results in subsection provides a method to determine impact through probabilistic objectives or constraints for linear and nonlinear systems. However, analyzing attack impact at a system-wide level becomes increasingly important. Thus, in the next subsection, we shift focus to impact metrics designed specifically for large-scale NCS.
%
%
%
\subsection{Impact metrics for large-scale systems}
\label{subsec:impact_large_scale}
For efficient analysis and design, large-scale control systems are typically divided into interconnected subsystems, often referred to as multi-agent systems \cite{olfati2007consensus}. A significant advantage of this approach is the ability to represent subsystem interconnections using graph theory, wherein the system is modeled as a graph. From a security perspective, such graph-theoretic models enable the quantification of security metrics via intuitive and scalable properties \cite{west2001introduction}, making them well-suited for large-scale systems. We illustrate this scalability advantage with the following example.

\begin{exmp}\label{label:Nordic}
\begin{figure}[!t]
    \centering
    \includegraphics[width=0.7\linewidth]{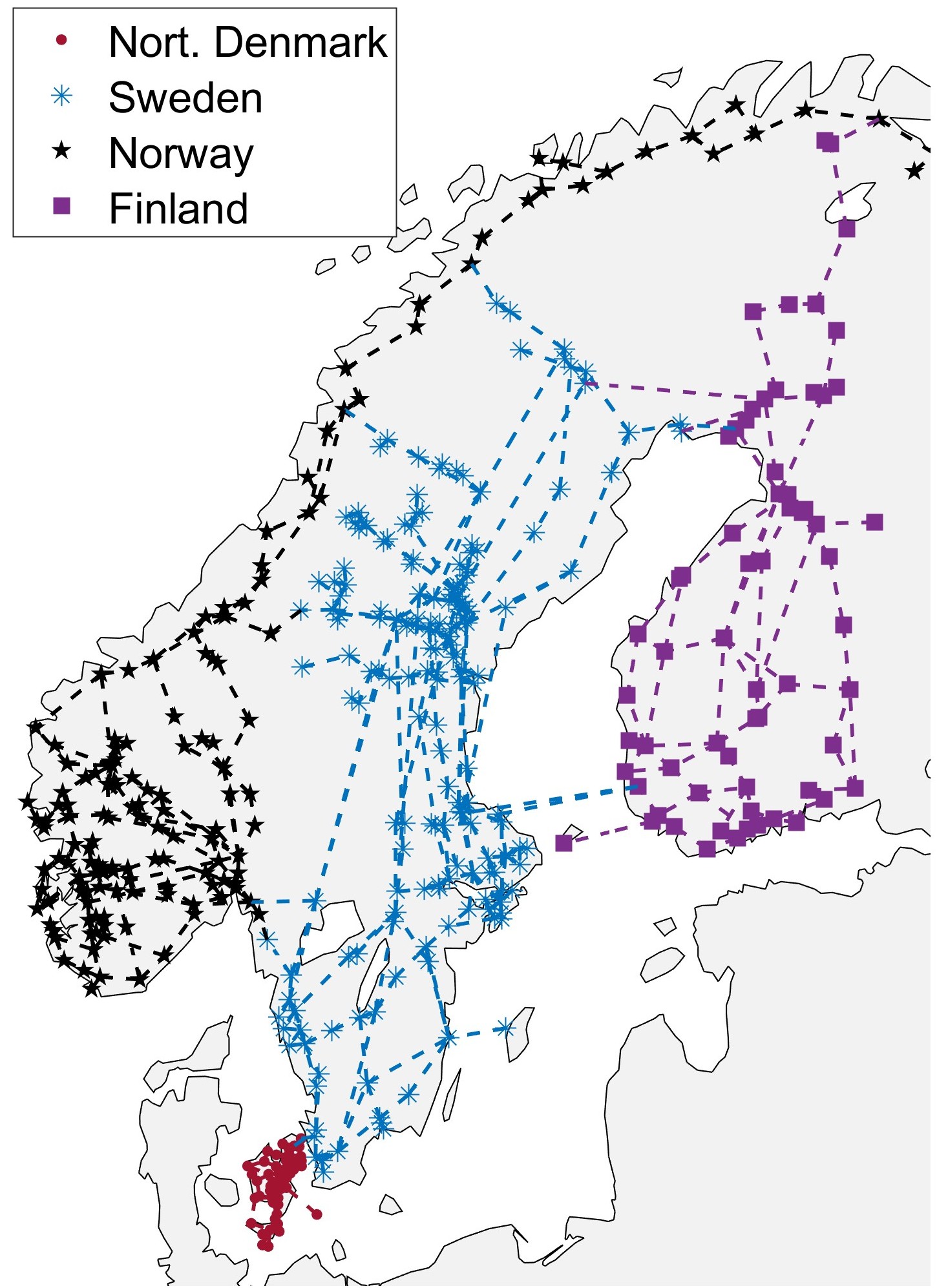}
    \caption{The Nordic490 power transmission grid comprises 498 buses (nodes) and 673 transmission lines (edges), spanning Sweden (blue), Northern Denmark (red), Norway (black), and Finland (purple), with connections outside the Nordic region excluded.}
    \label{fig:N490}
\end{figure}
Consider the large-scale Nordic power transmission grid, Nordic490, introduced in \cite{kumar2021open} and illustrated in Figure~\ref{fig:N490}. The power injected at each node (bus) is governed by the nonlinear swing equations \cite[Chapter 8.1.2]{tegling2018fundamental}. Assuming the grid operates near a stable equilibrium, the swing equations can be linearized into second-order differential equations. Thus, the dynamics of the Nordic grid can be described by a graph with \(N = 498\) nodes, where each node exhibits second-order dynamics. The structure of the grid is represented by a graph in which edges correspond to the interconnections between buses. Thus, we can model the Nordic power transmission grid as a multi-agent system. \hfill \(\triangleleft\)
\end{exmp}

One of the earliest applications of the multi-agent system approach to security is presented in \cite{riehl2017centrality}, where the impact metric is defined using the betweenness centrality \cite{freeman1977set} of the graph representing the control system. While no specific detection mechanism is introduced, the model assumes that the malicious impact is zero if the attacker and defender target the same node, irrespective of its centrality.
As a result, given an attack node $a$ and defender node $d$, the impact metric $\mathcal{J}$ in \eqref{eq:impact} is formulated as follows
\begin{align}
    \mathcal{J} \triangleq \mathcal{C}_b(a) \times \big( \, 1 - \mathbb{I}_{d}(a) \, \big), \label{impact_betweenness_centrality}
\end{align}
where $\mathcal{C}_b(a)$ is the betweenness centrality measure of node $a$ and $\mathbb{I}_d(a)$ is an indicator function that takes $1$ if $a \equiv d$ and $0$ otherwise. However, \cite{riehl2017centrality} does not account for system dynamics, and thus no detection model $f_D(r)$ is specified. A similar graph-theoretic approach without dynamics is applied in \cite{forsberg2023power} to evaluate vulnerabilities in the Nordic power system topology.

To address the omission of system dynamics in \cite{riehl2017centrality,forsberg2023power}, the study \cite{pirani2021game} introduces an impact metric leveraging the grounded Laplacian matrix, denoted as $L_g$ ($A$ in \eqref{P} replaced with $-L_g$ ). A grounded Laplacian matrix is obtained by removing rows and columns corresponding to some nodes called grounded nodes \cite{pirani2015smallest}. The positivity of $-L_g$ allows for quantifying worst-case energy transfer from attack nodes to performance nodes using the inverse grounded Laplacian matrix. 

In order to help us define the impact matrix, we introduce some notation. Let us consider a multi-agent system where the augmented state of the network is denoted by $x$. When there is an attack input, let $E_a$ represent the attack input matrix. Let there be an agent whose states, denoted by $y_m$, are monitored for attack. Similarly, let there be an agent whose states, denoted by $y_m$, represent the performance of the entire network. In other words, the closed loop from the attack input to the performance and monitor output can be given as 
\begin{align}
\dot{x} &= A x + E_a a,\;y_p = C_px,\;y_m = Cx.
\end{align}
Using the notations above, \cite{pirani2021game} define the impact metric as \eqref{eq:impact} where
\begin{align}
    \mathcal{J} \triangleq \lambda \, \bar \sigma \big( C_p^\top L_g^{-1} E_a \big ) - \bar \sigma \big( C^\top L_g^{-1} E_a \big ),  \label{pirani_impact_metric}
\end{align}
where $\lambda > 0$ is the trade-off between the performance loss and the detection energy, similar to \cite{shames2017security}. In the grounded Laplacian dynamics \cite{pirani2021game}, the largest singular value $\bar \sigma(\cdot)$ is equivalent to the DC-gain \cite{farina2011positive}.

It is worth noting that instead of using stealthiness condition ($f_D(r) < \tau$ in \eqref{eq:impact}), the impact metric \eqref{pirani_impact_metric} incorporates the malicious impact and detection impact in one equation, enabling the security analysis to solely rely on the inverse of grounded Laplacian matrix. This inverse is also used to compute the trace of the observability Gramian, which measures the average malicious impact of attack nodes on the entire network \cite{pirani2020graph}.

With the same spirit as \eqref{pirani_impact_metric}, the work \cite{pirani2021strategic} formulates the impact metric in terms of reachability/controllability set with attack nodes as inputs. More specifically, the impact metric for a set of attack nodes $\mathcal{A}$ in \eqref{eq:impact} is formulated as follows
\begin{align}
    \mathcal{J} \triangleq \lambda \, \vert \mathcal{C}^r(\mathcal{A})\vert - \big( \text{gnr}(R(\mathcal{A})) - n \big), \label{pirani_metric_reachable}
\end{align}
where $\lambda > 0$ is the trade-off between the performance loss and the detection energy, $n$ is the number of nodes in the graph describing the system, $\mathcal{C}^r(\mathcal{A})$ is the set of nodes that can be reached from a node in $\mathcal{A}$, $\text{gnr}(R(\mathcal{A}))$ is the generic rank of the following Rosenbrock pencil matrix with inputs from nodes in $\mathcal{A}$
\begin{align*}
    R(\mathcal{A}) \triangleq \left[ \begin{array}{cc}
       A - zI  & E_a \\
       C  &  0
    \end{array} \right],~~ \text{for almost all}~ z \in \mathbb{C}.
\end{align*}
Given the structural metric \eqref{pirani_metric_reachable}, one can leverage results from structural system theory \cite{ramos2022overview} to obtain well-established solutions to security problems without introducing new methods. 

Another approach incorporating the system dynamics into the impact \eqref{impact_betweenness_centrality} studies the fundamental limitations of stealthy attack policies \cite{shinohara2024optimalatm,shinohara2024optimaltac,weerakkody2016graph,zhang2023structural}, which guarantee the stealthiness condition $f_D(r) < \tau$ for classical anomaly detection mechanisms (see Table~\ref{tab:detectors}).
These studies formulate the impact metric \eqref{eq:impact} as follows
\begin{align}
    \mathcal{J} \triangleq I_{\infty} \times \big( 1 - \mathbb{I}_{\mathcal{F}}(\mathcal{A}) \big), \label{impact_fundamental}
\end{align}
where $I_{\infty}$ is the unbounded malicious impact and $\mathcal{F}$ is the set of nodes with some fundamental limitations, i.e., if the set of attack nodes $\mathcal{A}$ is contained by $\mathcal{F}$, then the impact  $\mathcal{J} = 0$, otherwise $\mathcal{J} = I_{\infty}$. Note that we can consider an unbounded malicious impact $I_\infty$ in theory but in practice where $I_\infty$ can be considered as a very high malicious impact. As a consequence, the metric \eqref{impact_fundamental} is binary, $0$ or $I_\infty$. The work  \cite{shinohara2024optimalatm,shinohara2024optimaltac} considers the sparse observability limitation, which was proposed in \cite{nakahira2018attack}. The perfectly undetectable limitation is considered in \cite{weerakkody2016graph,zhang2023structural}.
 
%
A straightforward approach in \cite{nguyen2023security} extends impact metrics originally designed for small-scale systems to large-scale systems by leveraging structural properties through a graph-theoretical representation. Specifically, the authors adopt the impact metric in \eqref{eq:OOG}, where $y_p$ denotes the outputs of performance nodes and $r$ corresponds to the outputs of monitor nodes. While this formulation remains valid for large-scale systems, evaluating the impact in \eqref{eq:OOG} becomes computationally intensive as system size increases.

To address this challenge, 
inspired by \eqref{impact_fundamental}, a graph-theoretical limitation in terms of dominating sets that yields infinite malicious impact is reported in \cite{nguyen2023security}. Building on this foundation, \cite{nguyen2024centrality} explores how centrality measures \cite[Chapter~3]{golbeck2013analyzing}, when combined with the impact metric in \eqref{eq:OOG}, can be used to guide the strategic placement of monitor nodes.

\subsection{Impact of covert attacks under uncertain process knowledge}\label{subsec:covert:impact}
\begin{figure}
    \centering
    \includegraphics[width=5cm]{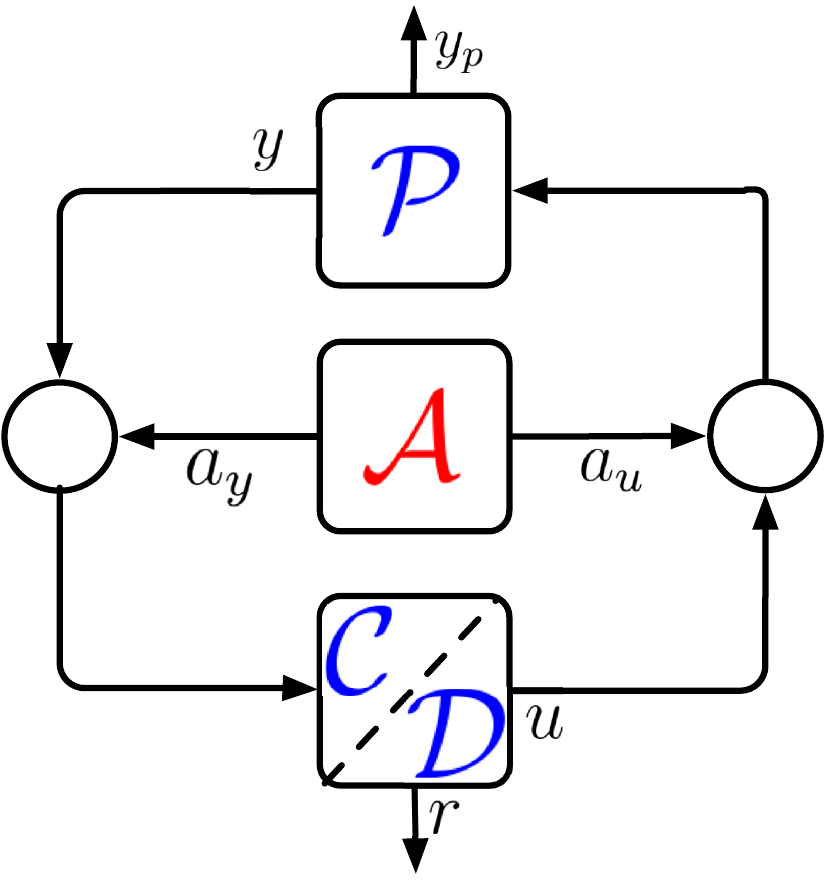}
    \caption{Networked control system under covert attack.}
    \label{fig:covert}
\end{figure}

In this section, we relax the Assumption \ref{ass:knowledge} and \ref{ass:covert}. In other words, we consider covert attacks \cite{smith2015covert} where both sensor and actuator are under attack. A covert attack policy can be described as follows (also see Figure~\ref{fig:covert}). 

Consider the attacker injecting a signal $a_u$ into the actuator channels. Since the adversary knows the process dynamics, the adversary computes the process output $a_y=Ga_u$ and injects the signal $-a_y$ into the sensor channels, thereby canceling the effect of the attack from the measurement sent to the detector. Here $G$ represents the transfer matrix of the process \eqref{P}. Since the effects of the attack signals do not reach the detector, covert attacks are hard to detect. 

In this section, we review works that quantify the impact of covert attacks. However, as mentioned before, the impact (performance loss) caused by covert attacks is unbounded when the adversary has perfect system knowledge. Considering imperfect adversarial knowledge of the process is critical to developing impact metrics for covert attacks. 

The paper \cite{muller2018risk} considers that the adversary injects an attack signal to change the reference signal followed by the plant. When the adversary has access to only the nominal model of the plant, but not the true model, the paper \cite{muller2018risk} quantifies the impact of covert attack as 
\begin{equation}\label{eq:Jez:H2}
\Vert G_{\tilde{r} \to e}\Vert_2^2 + \Vert G_{\omega \to e}\Vert_2^2 + \Vert G_{r \to e}\Vert_2^2
\end{equation}
where $G_{\tilde{r} \to e}, G_{\omega \to e}$ and $G_{r \to e}$ are the transfer matrices from the reference trajectory of the adversary, the disturbance signal and the reference signal to the tracking error, respectively. Thus, the metric \eqref{eq:Jez:H2} captures the sum of the $\mathcal{H}_2$ norm of the transfer matrices. However, the metric does not consider any detection constraints.

\begin{figure}
    \centering
    \includegraphics[width=7cm]{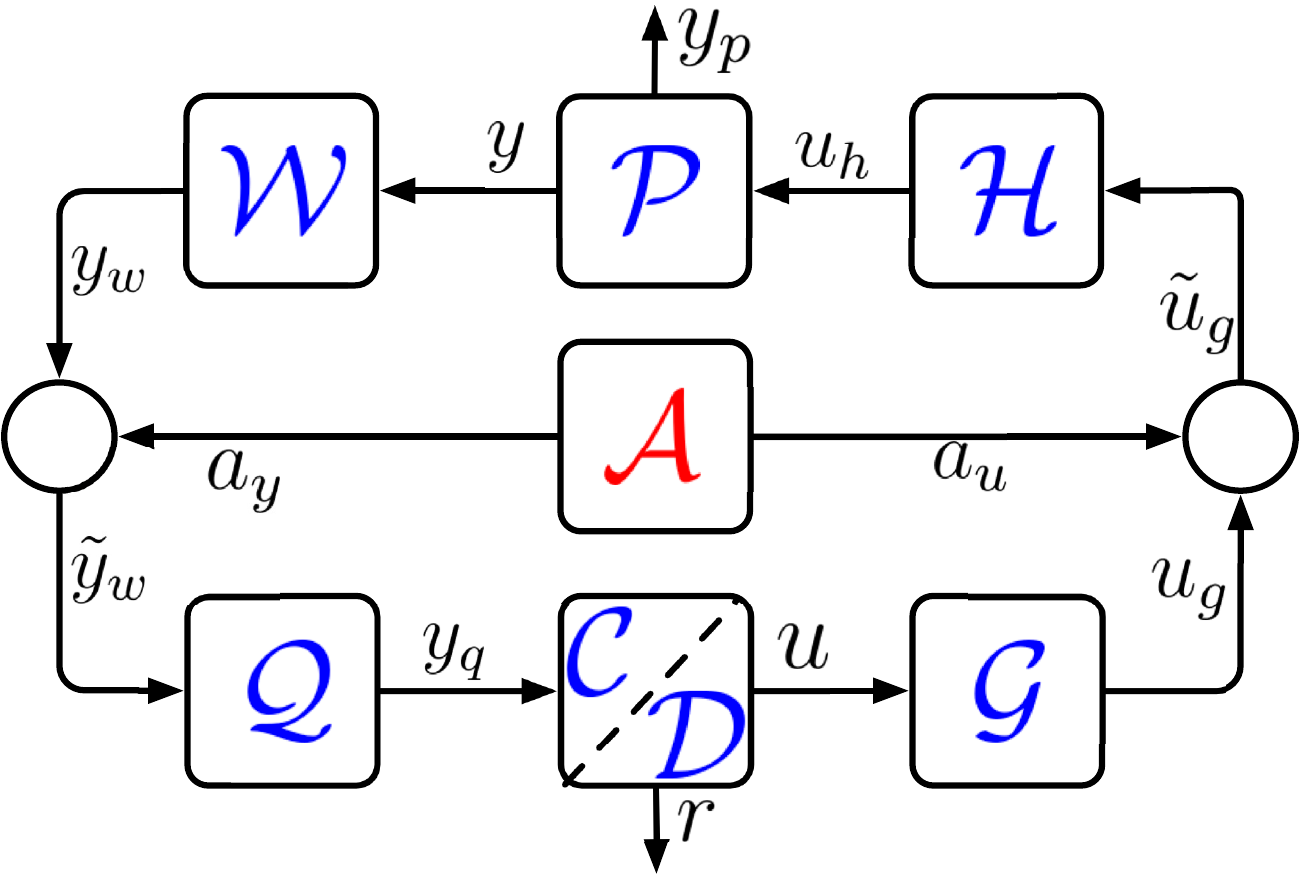}
    \caption{Networked control system with multiplicative watermarks under covert attack.}
    \label{fig:covert:mwm}
\end{figure}

The work \cite{gallo2021design} considers a multiplicative watermark to detect covert attacks. The concept of multiplicative watermarking can be explained as follows using Figure~\ref{fig:covert:mwm}. Consider that the plant output is fed to an invertible linear system $\mathcal{W}$, the output of which is sent over the wireless network. The signal received by the controller is again fed through a linear system $\mathcal{Q}$, the output of which is used to estimate the plant states. If $\mathcal{W}=\mathcal{Q}^{-1}$, then the plant states are estimated accurately in the absence of attacks. 

The paper \cite{gallo2021design} considers an adversary that does not have knowledge about the filter dynamics $\mathcal{W}$. Then, the authors use OOG, defined in \eqref{eq:OOG}, to study the impact of covert attacks. The AEC-OOG has also been used to study the impact of covert attacks in \cite{gallo2024switching}.

In summary, in this section we reviewed the literature on methods to quantify the impact of FDI attacks. An overview of the methods discussed in this section is provided in Table~\ref{tab:review:impact}. In the next section, we provide an overview of the methods that quantify the resource needed by a stealthy adversary. 
\begin{table*}
\caption{Summary of impact metrics discussed in section \ref{sec:impact}. Here, SDP represents a Semi-definite Program, QLCP represents a Quadratically constrained Linear Program, QCQP represents a Quadratically Constrained Quadratic Program, LP represents a Linear Program, MDP represents Markov Decision Process, and BMI represents a Bi-linear Matrix Inequality.}
\label{tab:review:impact}
\vspace{-5pt}
\begin{center}
\begin{tabular}{||p {3.2cm}| c | p{2.2cm} | c | p{1.9cm} | p{3.3cm}||} 
 \hline
$\mathcal{J}$ in \eqref{eq:impact} & $f_D(r)$ in \eqref{eq:impact} & Constraints on $a_u/a_y$ & Paper & Problem structure &Remarks \\ [0.5ex] 
 \hline\hline
 $\lim\limits_{k \to \infty} \vert \mathbb{E} [x[k]] \vert $ & CUSUM & None & \cite{murguia2016characterization} & SDP & \;\\
\hline
Volume$\left(\mathcal{R}_e \right)$ & $\chi^2$ &None & \cite{murguia2020security} & SDP & $\mathcal{R}_e$ is defined in \eqref{eq:Re}\\
\; & \; & None & \cite{hashemi2018comparison} & SDP & Geometric approach\\
\hline 
$\left( \min. \text{dist} [\mathcal{R}_x, \mathcal{S}_x] \right)^{-1}$ & $\chi^2$ & None &\cite{murguia2020security} &SDP & $\mathbb{R}_x$ and $\mathcal{S}_x$ defined in \eqref{eq:reachset}, \eqref{eq:safeset}\\
\hline
$\lim\limits_{k \to \infty} \Vert e[k]] \Vert_2 $ & KL detector & None & \cite{sui2020vulnerability} & Closed-form & \;\\
\hline
$\left( \mathbb{E}\left[ \sum\limits_{k=0}^{N} \Vert x[k]-x^* \Vert_{Q}\right] \right)^{-1}$ & $\Vert \epsilon[k] \Vert_2 \leq \tau, \forall k $ & None & \cite{chen2017optimal} & Closed-form & $\begin{aligned} x^* &\text{- target state}\\ \epsilon &\text{- Innovation}\end{aligned}$\\\hline
$\lim\limits_{k \to \infty} \vert \mathbb{E} [e^2[k]] \vert $ & Generic & None & \cite{bai2014kalman} & Closed-form & Scalar system\\
\hline 
$\lim\limits_{k \to \infty} \vert \mathbb{E} [e[k]] \vert $ & None & $\vert a_y[k]\vert \leq \bar{a},\;\forall k$ & \cite{sargolzaei2021secure} & Closed-form & Non-linear input-affine process\\
\hline
$\Vert y_p[k_0:k_0+N]\Vert_p$ & $\ell_q$ detector & None & \cite{teixeira2013quantifying} & Closed-form, SDP & \;\\
\hline
$\Vert y_p[k_0:k_0+N]\Vert_\infty$ & None & $ \vert {u}_a[k] \vert \leq \bar{u}, \;\forall k$ & \cite{milovsevic2017exploiting,milovsevic2018quantifying} & LP & $u_a$ is the control signal under attack. \\
\hline
$\Vert x[k]\Vert$ & KL detector & None & \cite{khazraei2022resiliency} & None & Nonlinear process   \\
\hline
$\Vert y_p[k_0:k_0+N]\Vert_\infty$ & $\ell_{\infty}$ detector & None & \cite{hirzallah2018computation} & LP & \;\\
\hline
$\Vert y_p\Vert_{\ell_2}$ & $\ell_2$ detector & None & \cite{teixeira2015strategic} & SDP &  \;\\
\hline 
$\Vert y_p\Vert_{\ell_2}$ & $\ell_2$ detector & $\lim\limits_{k \to \infty} x[k] = 0$ & \cite{teixeira2019optimal,anand2023risktac} & SDP &  Extensions for uncertain systems\\
\hline 
$\Vert y_p\Vert_{\ell_2}$ & $\ell_2$ detector & $\Vert a_u\Vert_{\ell_2} \leq \epsilon_a$ & \cite{anand2023risk} & SDP &  \;\\
\hline 
$\Vert y_p\Vert_{\ell_1}$ & $\ell_1$ detector & $\Vert a_u\Vert_{\ell_1} \leq \epsilon_a$ & \cite{anand2024scalable} & LP &  Positive systems\\
\hline 
$\lim\limits_{k \to \infty} \vert \mathbb{E}[e[k]]\vert$ & FAR $\leq \alpha$ & None & \cite{milovsevic2017exploiting} & QCQP &  \;\\
\hline 
$\mathbb{E}[\Vert y_p[k_0:k_0+N]\Vert_{\infty}]$ & KL detector & None & \cite{milovsevic2019estimating} & Iterative algorithm &  \;\\
\hline 
$\mathbb{E}[\Vert e[k]\Vert_{\infty}]$ & None & $\vert a_u[k]\vert \leq \bar{a} \;\forall k$ & \cite{wang2017security} & Closed-form &  Process with sector non-linearities\\
\hline 
$\mathbb{E}\left[\sum_{k=0}^{N} R(k)\right]$ & FAR $\leq \alpha$ &None & \cite{sasahara2022attack} & LP &  MDP, $R(\cdot)\;$- reward\\
\hline
$\mathbb{P}[\vert y_p[k_0:k_0+N]\vert >1]$ & KL detector &None & \cite{milovsevic2019estimating} & Iterative algorithm &  Yields an upper bound\\
\hline
$\mathbb{P}[\vert y_p[k_0:k_0+N]\vert >1]$ & None & $\vert a_u[k]\vert \leq \bar{a} \;\forall k$ & \cite{ding2016security} & Nonlinear sufficient conditions &  Nonlinear control-affine process\\
\hline
$\mathcal{C}_b(a) \times \big( \, 1 - \mathbb{I}_{d}(a) \, \big)$ & None & None & \cite{riehl2017centrality} & Greedy search & $\mathcal{C}_b(\cdot)$ is a centrality measure \\ 
\hline
$\begin{aligned}
&\lambda \, \bar \sigma \big( C_p^\top L_g^{-1} E_a \big ) \;- \\
&\bar \sigma \big( C^\top L_g^{-1} E_a \big )
\end{aligned}$  & None  & $\Vert a\Vert_{0} \leq \bar a$ & \cite{pirani2021game} & Closed-form & $\bar a$ - maximum number of attack nodes, \\
\cline{1-5}
$\begin{aligned}
&\lambda \, |\mathcal{C}^r(\mathcal{A})|\;-\\
&\big( \text{gnr}(R(\mathcal{A})) - n \big)
\end{aligned}$ & None & $\Vert a\Vert_{0} \leq \bar a$ & \cite{pirani2021strategic} & Closed-form & $\lambda $ - trade-off parameter, gnr$(\cdot)$ - generic rank \\ 
\hline
$\begin{aligned}
&\Vert G_{\tilde{r} \to e}\Vert_2^2 + \Vert G_{\omega \to e}\Vert_2^2 +\\
& \Vert G_{r \to e}\Vert_2^2
\end{aligned}$  & None  & White noise & \cite{muller2018risk} & QLCP & Covert attack \\
\hline
$\Vert y_p \Vert_{\ell_2}^2$ & $\ell_2$ detector  & None & \cite{gallo2021design} & BMI & Covert attack \\
\hline
\end{tabular}
\end{center}
\end{table*} 

%% file: Resource.tex
%
\begin{table*}
\caption{Summary of resource metrics.}
\label{tab:review:security}
\begin{center}
\begin{tabular}{||p{2.1cm} | p{2.2cm} | p{1.5cm} | c | p{3cm} | p{3cm}||} 
 \hline
Objective in \eqref{eq:security} & Constraint in \eqref{eq:security} & Constraints on $a_u/a_y$ & Paper & Problem structure &Remarks \\ [0.5ex] 
 \hline\hline
$\Vert a_y\Vert_0$ & $0 \neq \tilde{y} \in \mathcal{B}$ & None & \cite{chong2016characterising} & LP & $
\begin{aligned}
&\mathcal{B}: \text{nominal outputs}\\
&\text{Autonomous system}
\end{aligned}$\\
\; & \; & \; & \cite{tang2018sensor} & Iterative algorithm & external inputs\\
\hline 
$\Bigg \Vert \begin{bmatrix} a_{u\backslash j}\\a_y \end{bmatrix} \Bigg \Vert_0$ & $r =0, a_j \neq 0$ & None & \cite{milovsevic2018security} & $s-t$ cut algorithm & Security of actuator $j$\\
\hline
$\Vert a\Vert_0$ & $(A,\,C_{-\Vert a \Vert_0})$ is unobservable & None & \cite{shinohara2024optimalatm,shinohara2024optimaltac} & Greedy algorithm & Economical sensor placement
\\ \hline
\end{tabular}
\end{center}
\end{table*} 
\subsection{Resource metrics for stealthy attacks}
The work \cite{chong2016characterising} considers an autonomous system where the sensor measurements are transmitted for monitoring over the network. Let $\mathcal{B}$ represent the set of nominal output trajectories of the autonomous system. Then \cite{chong2016characterising} defines {a} resource metric as the number of sensors that need to be corrupted to ensure that the attacked output trajectory $\tilde{y}$ stays within $\mathcal{B}$, i.e., $\tilde{y} \in \mathcal{B}$. In other words, \cite{chong2016characterising} defines the resource metric as 
\begin{equation}\label{eq:michelle1}
\inf_{0 \neq \tilde{y} \in \mathcal{B}} \Vert a \Vert_0,
\end{equation}
{where $\Vert a \Vert_0$ indicates the number of non-zero entries of $a$.}
Thus the metric in \eqref{eq:michelle1} is equivalent to the optimization problem \eqref{eq:security} when $p=0$ and the detection constraint is replaced by $\tilde{y} \in \mathcal{B}$. The paper \cite{chong2016characterising} also describes five different methods to efficiently compute the resource metric \eqref{eq:michelle1}. The work was extended to systems with external inputs in the noise-free case in \cite{tang2018sensor} and kernel representations in \cite{tang2019linear}.

The $H_{\infty}$ metric \cite{zames1981feedback} is a classical metric used to quantify the performance degradation caused by disturbances. The $H_{\infty}$ metric was modified and used to quantify security in \cite{bopardikar2016h}. To detail, the paper \cite{bopardikar2016h} considers that both the sensors and actuators are under attack. Then \cite{bopardikar2016h} quantifies security as the maximum energy of the attack input for which the attack is stealthy. In other words, the resource metric is defined as 
\begin{equation}\label{eq:Hinf2}
\sup_{a} \left\{ \; \Vert a \Vert_{\ell_2}^2\;\big\vert\; \Vert r \Vert_{\ell_2}^2 \leq \tau\right\}
\end{equation}
The metric \eqref{eq:Hinf2} is shown to be equivalent to a convex SDP. Such energy metrics have also been used to analyze the security of power grids (see Section~\ref{sec:application}).

As \eqref{eq:michelle1} postulates, maintaining stealthiness only when the sensors are under attack is quite hard. In reality, the adversary can also gain access to some of the actuators. Thus, the paper \cite{milovsevic2018security} postulates a resource metric of an actuator (say $A_i$) as the minimum number of sensors and actuators that needs to be compromised in addition to $A_i$, such that a perfectly undetectable attack is possible. Here, an attack is defined as perfectly undetectable when $r=0$ is in the presence of attacks. In other words, \cite{milovsevic2018security} defines the security metric as the value of the optimization problem 
\begin{equation}\label{eq:jez:Allerton}
\inf_{0 \neq a} \left\{ \; \Vert a \Vert_0\;\big\vert \;r =0\right\}
\end{equation}
whose value can be found using algebraic conditions \cite[Theorem 1]{milovsevic2018security} for small scale systems. 

It is shown in \cite{milovsevic2020actuator} that the optimization problem \eqref{eq:jez:Allerton} is \emph{NP-hard} for large-scale systems. Thus, \cite[Section IV]{milovsevic2020actuator} proposes an upper bound of the resource metric \eqref{eq:jez:Allerton} which can be computed using polynomial time $s-t$ cut algorithm \cite{stoer1997simple}. The upper bound of the resource metric can be efficiently computed for large-scale systems. The authors also provide sufficient conditions for the upper bound to exist.

Another issue with the security metric in \eqref{eq:jez:Allerton} is \emph{fragility}. Consider a large-scale system where the system parameters change over time. When the parameters change, the security metric \eqref{eq:jez:Allerton} changes drastically \cite[Example 2]{milovsevic2020actuator}. Thus, if the system is secure (acceptable value of the metric) with one set of parameters, the system might not be secure when the parameter changes. For instance, if a power grid is secure in one configuration, it might not be secure when a microgrid detaches. Thus, using structural system framework \cite{lin1974structural}, the work \cite{gracy2021security} develops a security index impervious to variations in system parameters. Next, we review metrics developed for multi-agent systems.
\subsection{Resource metrics for multi-agent systems}
As described in Section~\ref{subsec:impact_large_scale}, multi-agent systems are described by graphs, and in general, we aim to provide graph-theoretic results/intuition/interpretation for security. The resource metric of perfectly undetectable attacks \eqref{eq:jez:Allerton} is considered in the context of multi-agent systems \cite{weerakkody2016graph,gracy2021security,zhai2022graph}. A notion of vertex separator to isolate two given nodes is introduced in \cite{weerakkody2016graph}, which is the backbone of a perfectly undetectable attack. Meanwhile, bipartite graphs describing input, output, and state nodes are utilized to provide algorithms for computing the resource metric \eqref{eq:jez:Allerton} in \cite{gracy2021security}.

In the problem of secure state estimation, a fundamental limitation on the sparse observability is introduced in \cite{nakahira2018attack}, i.e., for $p$ attacked sensor nodes, $2p$-sparse observability is needed to provide a resilient state estimate. This fundamental limitation implies that a system remains observable after removing arbitrary $2p$ sensor nodes. As a consequence, the following resource metric can be considered
\begin{equation}\label{eq:2pobservability}
\begin{aligned}
\inf_{a} & \quad  \Vert a \Vert_0\\
\text{s.t.} &\quad (A,\,C_{-\Vert a \Vert_0}) ~ \text{is unobservable},
\end{aligned}
\end{equation}
where $C_{-\Vert a \Vert_0}$ is the measurement matrix $C$ in \eqref{P} after removing arbitrary $\Vert a \Vert_0$ rows. The problem \eqref{eq:2pobservability} is NP-hard in general. However, multi-agent systems described by graphs enable the work \cite{shinohara2024optimalatm,shinohara2024optimaltac} to show that \eqref{eq:2pobservability} is coNP-complete (see \cite{goldreich1999introduction} for the definition of NP-hard and coNP-complete). Finally, we note that if a solution to the optimization problem \eqref{eq:2pobservability} does not exist, then it guarantees that there exists a resilient state estimator. To address the fundamental limitation of $2p$-sparse observability and to enhance the system resilience, the authors in \cite{mitra2021impacts} propose a selection algorithm for protecting several sensor nodes to increase the adversarial resource metrics \eqref{eq:2pobservability}.

The work \cite{taheri2020undetectable} develops conditions under which an attack is undetectable. It is shown if an agent that is the root of a rooted spanning tree is under a cyber attack, the attack is undetectable by any agent in the entire network (see \cite{wilson1979introduction} for the definition of a rooted spanning tree). Thus, while studying security, the probability of such nodes being attacked must be high. 
\subsection{Resilience metrics}\label{subsec:resilience}
A different approach to quantifying security is to consider resilience \cite{segovia2020cyber}. Previously, we defined security as the resources the adversary needs to conduct a stealthy attack. Resilience, on the other hand, quantifies the ability of an NCS to withstand the adversary, and the ability of the system to recover from a severe attack. 

For example, \cite{de2021resilient,gong2023resilient} define resilience as the maximum duration a DoS attack can persist while still maintaining the stability of the closed-loop system. This allowable duration is typically characterized using a semidefinite program (SDP). Similar frameworks have been explored in \cite{de2016networked,perodou2021critical}. In \cite{zhang2023decentralized}, the authors consider asynchronous state updates, while \cite{griffioen2024ensuring} emphasizes safety guarantees. In \cite{griffioen2024ensuring}, the notion of a safe set is defined similar to \eqref{eq:safeset}.

The work \cite{huang2018assessing} first computes the probability that a given sensor or actuator is under attack (similar to resource metrics) using a Bayesian network. Using these probabilities as inputs, \cite[Section V.D.]{huang2018assessing} quantifies resilience as the amount of time the NCS can be under attack before the safety constraint is violated.

The article \cite{sandberg2021stealthy} defines resilience as the ratio of time the attack signal is visible in the detection output $y_r$ to the amount of time the effect of attack signals is visible in the performance output. The attack is said to be visible in the signal $y_r$ if $y_r \geq \delta$ for a given $\delta$. The work provides algebraic conditions to determine if the system is vulnerable to attack. 

The work \cite{longo2017approach} defines resilience using the characteristics of the closed-loop step response. In particular, when a change (such as attacks) affects the system, \cite{longo2017approach} quantifies resilience using the following metrics
\begin{enumerate}
    \item \textbf{Settling time:} Time between attack onset and the time the states reach their nominal values.
    \item \textbf{Peak overshoot:} The maximum state deviation caused from the nominal state values (in $\%$).
    \item \textbf{Peak time:} The time taken to cause the maximum state deviation. 
    \item $\gamma$-\textbf{percentile time:} The time taken to reach $\gamma \%$ of the peak overshoot.  
\end{enumerate}

In the context of multi-agent systems, \cite{sun2019quantifying} defines resilience in terms of structural controllability. Specifically, the work defines resilience as the maximum number of edges that can be removed while maintaining network controllability. In contrast to analytical approaches, the work \cite{dhiman2021using} provides a learning-based approach to quantify resilience for multi-agent systems. 

%% file: Risk.tex
\begin{figure}
\centering
\includegraphics[width=8cm]{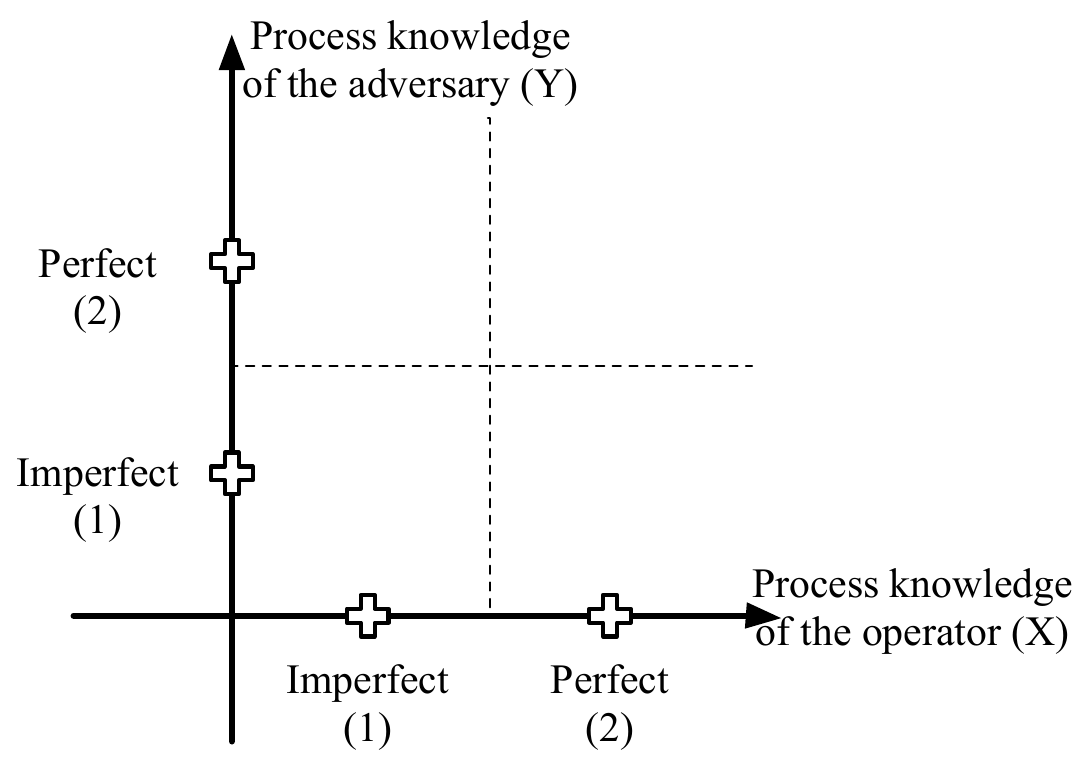}
\caption{Different information structure for the operator and the adversary (Figure adapted from \cite{coimbatore2024risk}). The works \cite{russo2021data,muller2018risk,anand2023risktac} consider the structure $(1,1)$,
the works \cite{harshbarger2020little,mukherjee2021secure,zhang2022design} consider the information structure $(2,1)$,
the work \cite{anand2022risk} consider the structure $(1,2)$, and the works \cite{hirzallah2018computation,murguia2020security} consider the structure $(2,2)$.}
\label{fig:knowledge}
\end{figure}
Until now, we reviewed methods from the literature to quantify security for an NCS with a given set of parameters. In other words, given the plant's parameters are constant and known, we defined methods to quantify security. However, the parameters of the plant need not be known accurately. Then, the risk metrics provide a way to quantify security {against cyber-attacks}. This is explained with an example next. 
\begin{exmp}[Impact metric]\label{exmp:risk:impact}
Consider an electric motor controlled over a network. The motor's time constant $\tau_c$ is not accurately known, i.e., $\tau_c = \tau +\delta,\,\delta \in \Omega$ where $\tau$ is the known nominal value and $\Omega$ is a discrete set. In other words, we consider that the matrices $A, B, C$ in \eqref{P} are uncertain (similar to adaptive control). When the value of $\delta$ is known to the operator and the adversary, then we can quantify the impact by the methods discussed in Section~\ref{sec:impact}. However, risk $R_m$ represents a measure of {the} performance loss over all possible time constant $\tau_c$. $\hfill \triangleleft$
\end{exmp}
Before we delve into the articles that quantify risk, it is important to ask the question
\begin{quote}
    If the parameters of the plant \eqref{P} are uncertain, as discussed in Example~\ref{exmp:risk:impact}, who is uncertain - the adversary or the operator?
\end{quote}

\begin{figure*}
    \centering
    \includegraphics[width=1.6\columnwidth]{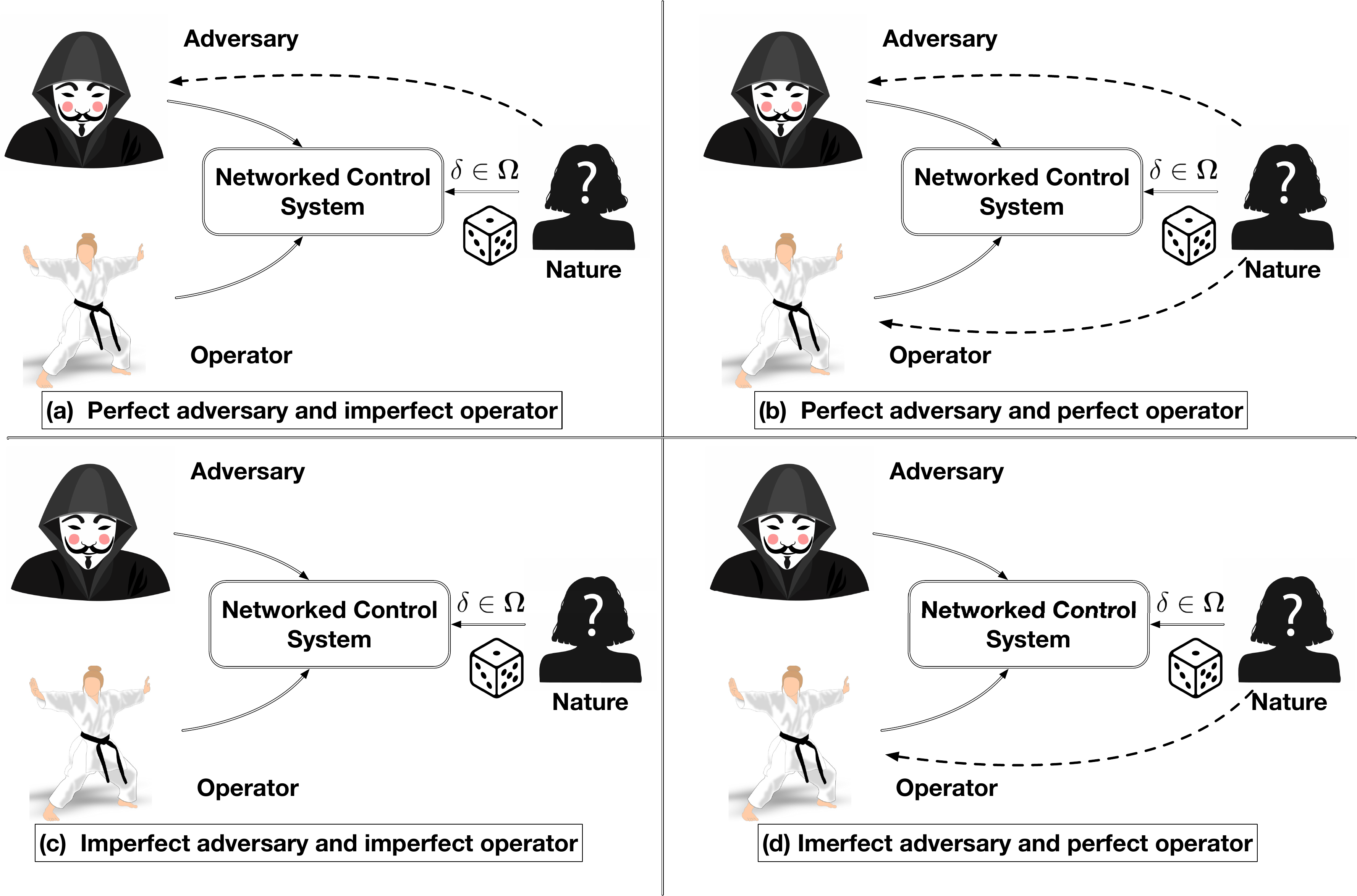}
    \caption{A pictorial representation of the information available to each player \cite{coimbatore2024risk}. Here, the players are the adversary, the operator, and nature. The dashed lines from player $A$ to player $B$ represent information from player $A$ available to player $B$. The solid lines represent the actions of the players. The nature is modeled as a player whose action $\delta \in \Omega$ is a random variable. The figure in the {top right} represents a setup where the realization of the uncertainty is known to both the adversary and the operator. The figure in the {top left} represents a setup where the realization of the uncertainty is known to the adversary but not the operator. In reality, it is hard for an adversary to know the true realization of the uncertainty. Thus, an omniscient adversarial setup is far from reality, but it can help us study a worst-case scenario. The figure in the {bottom left (right)} represents a setup where the realization of the uncertainty is unknown to the adversary and (known) to the operator.}
    \label{fig:layout}
\end{figure*}

To answer the above questions, consider two players: the adversary and the operator. We define a player as perfect (imperfect) when s/he knows the process parameters accurately (with some uncertainty). Then, different information structures arise, as depicted in Figure~\ref{fig:knowledge} and Figure~\ref{fig:layout}, with some comments in Table~\ref{tab:information}.
\begin{remark}
The definition of risk measures used in this section \cite{delbaen2000coherent} is different from the one in \cite{teixeira2015secure_mag}. In \cite{teixeira2015secure_mag}, the authors define risk as the tuple \emph{Risk {$\triangleq$} \{Scenario, Impact, Likelihood\}} where the impact and likelihood are quantified by methods similar to the ones described in Section~\ref{sec:impact} and \ref{sec:security}, respectively. $\hfill \triangleleft$
\end{remark}
\begin{table}[]
    \caption{Comment on various information structures. Here, \textbf{A} represents an adversary, and \textbf{O} represents an operator. Since the operator might have some process knowledge available, we can consider $(1,1)$ as a realistic setup. If we consider that the adversary has less knowledge about the process dynamics than the operator $(2,1)$, then it is an optimistic setup. On the other hand, if the adversary has complete system knowledge, we consider it to be the worst-case, i.e., $(1,2)$ and $(2,2)$.}
    \label{tab:information}
    \centering
    \begin{tabular}{||c|c||}
    \hline
    Structure & Remark\\
    \hline 
    $(1,1)$ & A realistic setup\\
    $(2,1)$ & An optimistic setup\\
    $(1,2)$ & A worst-case \textbf{A} and a realistic \textbf{O}\\
    $(2,2)$ & A worst-case \textbf{A} and an optimistic \textbf{O}\\
        \hline
    \end{tabular}
\end{table}
\subsection{Perfect knowledge for both players} 
In this scenario, the adversary and the operator have perfect process knowledge. Although such a setup is far from reality, most works in the literature consider such a setup. The advantage of this setup is that it helps us analyze the worst-case impact and develop a worst-case mitigation strategy. When both players have perfect information, the value of the impact/resource metrics can be determined as mentioned in Section~\ref{sec:impact} and Section~\ref{sec:security}, respectively. Here, there is no need for risk metrics.
\subsection{Imperfect operator and perfect adversary} 

\begin{figure*}
    \begin{minipage}[b]{0.48\linewidth}
        \centering
        \includegraphics[width=\linewidth]{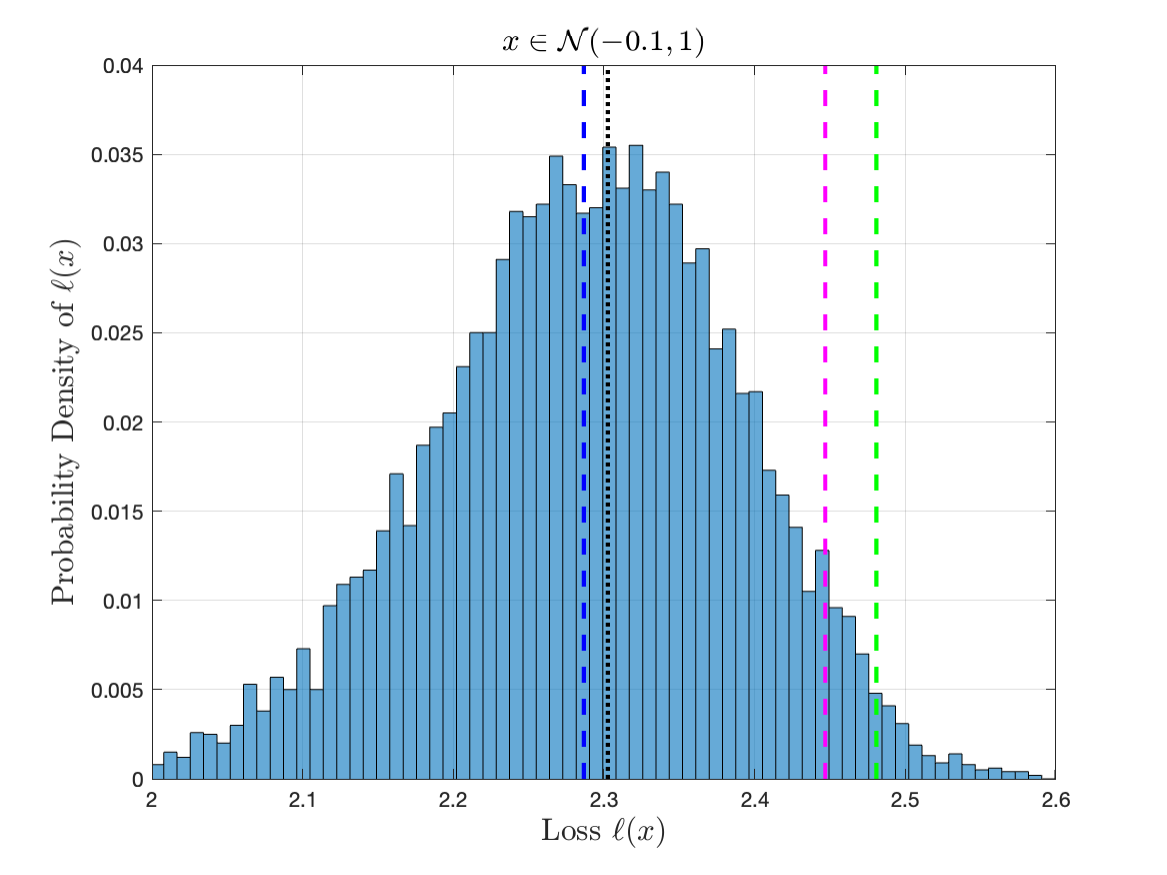}
    \end{minipage}
    \hfill
    \begin{minipage}[b]{0.48\linewidth}
        \centering
        \includegraphics[width=\linewidth]{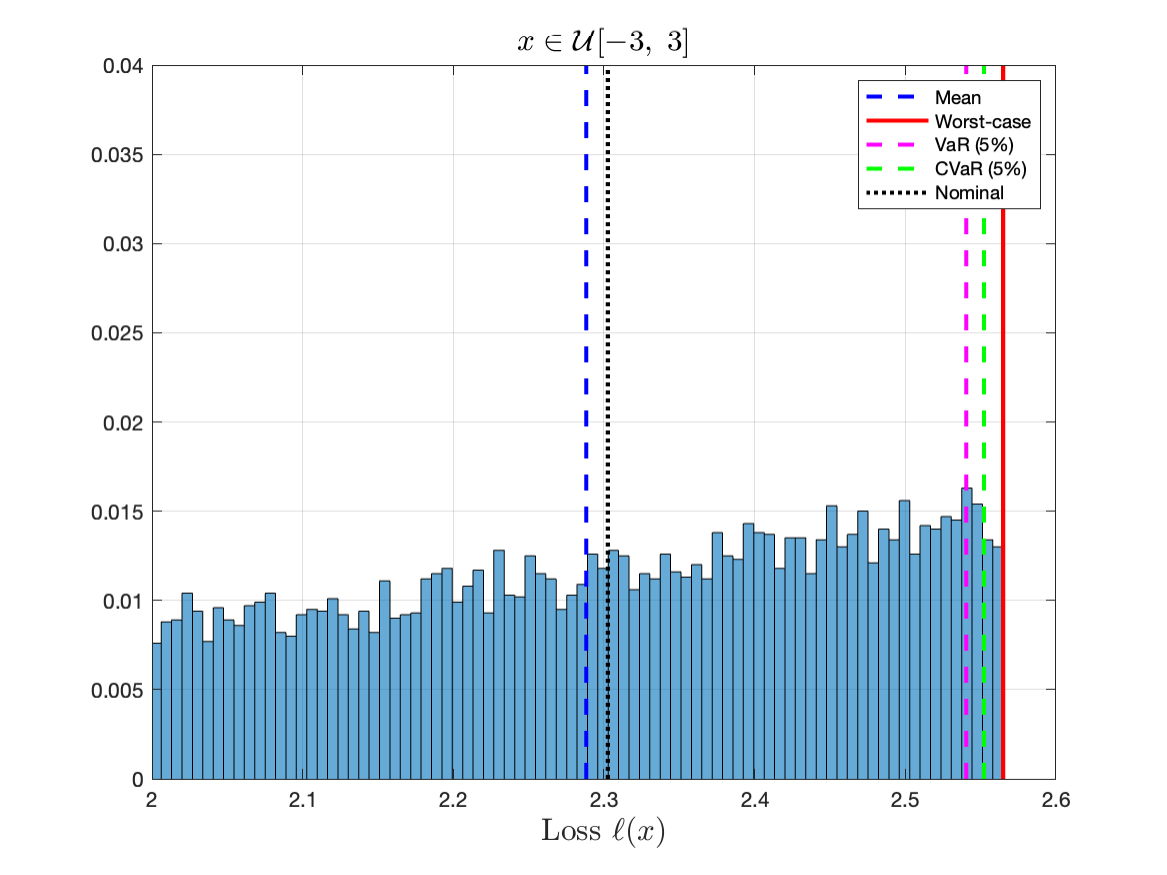}
    \end{minipage}
    \caption{Pictorial representation of probabilistic risk metrics for a loss function \(\ell(x): \mathbb{R} \to \mathbb{R}\), where \(x\) is a discrete random variable distributed normally (left) and uniformly (right). The expected loss is the mean loss value given by \( \mathbb{E}[\ell] = \sum_{x} \ell(x)\, p[\ell(x)]\), where \(p[\ell(x)]\) denotes the probability mass function of \(\ell(x)\). The worst-case loss is the maximum value of the loss function given by \(\sup_{x} \ell(x)\). For any given \(\alpha \in (0,1)\), the Value-at-Risk (VaR) at level \(\alpha\) is defined as \(\mathrm{VaR}_{\alpha}(\ell) = \inf \left\{ \bar{\ell} \in \mathbb{R} \mid \mathbb{P}[\ell(x) \leq \bar{\ell}] \geq 1 - \alpha \right\}\). Thus, \(\mathrm{VaR}_{\alpha}\) is the smallest threshold \(\bar{\ell}\) such that, with probability at least \(1 - \alpha\), the loss \(\ell(x)\) does not exceed \(\bar{\ell}\). Similarly, the Conditional Value-at-Risk (CVaR) at level \(\alpha\) is defined as the average of the loss values exceeding \(\mathrm{VaR}_{\alpha}\): 
    $\mathbb{E}[\ell(x)\vert \ell(x) \geq \mathrm{VaR}_{\alpha}(\ell)]$. Finally, the nominal loss is defined as the loss under nominal conditions: \(\mathbb{NL}(\ell) = \ell(0)\).
    }
    \label{fig:risk}
\end{figure*}

Consider an imperfect operator as described in Example \ref{exmp:risk:impact}. Let the set of all models consistent with process data be represented by $\mathcal{P}$. We assume the set $\mathcal{P}$ to be finite (see remark \ref{rem:scenario}) and known to the operator. Let us denote the true process model as ${P}_t$. We consider that the adversary knows $P_t$ (a worst-case setup) and injects an attack by solving the following optimization problem 
\begin{equation}
I_m(P_t) \triangleq \left\{\begin{aligned}
\sup_{a} & \quad \mathcal{J}_t\\
\text{s.t.}& \quad f_D(r_t) \leq \tau
\end{aligned}\right.,
\end{equation}
where $\mathcal{J}_t$ represent the performance loss of the true plant, and $r_t$ is the residual signal generated by the true plant. Since the operator only knows the distribution \(\mathcal{P}\), the exact performance loss caused by the adversary on the true plant cannot be determined. However, the expected performance loss can be computed as
\begin{equation}\label{eq:E}
\mathbb{E}_{P \sim \mathcal{P}}[I_m(P)] = \sum_{i=1}^N I_m(P_i)\, p(P_i),
\end{equation}
where \(\{P_i\}_{i=1}^N\) are the possible realizations of the plant, \(p(P_i)\) is the probability mass function of \(\mathcal{P}\), and \(N = |\mathcal{P}|\).

In \eqref{eq:E}, the impact $I_m(\cdot)$ is a function of the random variable $P_i$, and thus the risk, which is a measure of performance
loss over all possible models $\mathcal{P}$, can be obtained using risk metrics. For instance, the paper \cite{anand2022risk} quantifies risk as follows
\begin{equation}
R_m \triangleq \displaystyle \mathcal{R}_{P \in \mathcal{P}} \left(I_m(P)\right),
\end{equation}
where $R_m$ denotes the value of risk measuring the performance loss over all possible $P \in \mathcal{P}$, and $\mathcal{R}$ represents a risk measure such as Value-at-Risk (VaR), Conditional VaR (CVaR), the expected loss (used in \eqref{eq:E}), nominal loss, or worst-vase loss. A pictorial representation of the various risk measures for a generic loss function is represented in Figure~\ref{fig:risk}.
Choosing a risk measure $\mathcal{R}$ is a design choice for the operator. For instance, the papers \cite{milovsevic2017exploiting,pan2018cyber,anand2023risktac} use the expected loss, \cite{anand2022risk} use the Value-at-Risk, \cite{muller2018risk,anand2022riskacc,anand2023risk,nguyen2022zero} use the CVaR, and \cite{muller2018risk,anand2023risk} uses the worst-case loss. 
\begin{remark}\label{rem:scenario}
If $\mathcal{P}$ is a continuous set, we can quantify the risk using a finite number of sample scenarios from $\mathcal{P}$. Probabilistic risk guarantees on the original set $\mathcal{P}$ can be given using scenario approach \cite{campi2018introduction}. $\hfill \triangleleft$
\end{remark}
%
\subsection{Imperfect adversary} 
Let us consider an imperfect adversary that only knows the set $\mathcal{P}$. For any given plant model $P \in \mathcal{P}$, let us define the following
\begin{equation}
    X(P,a) \;{\triangleq}\; \mathcal{J}(P,a) \times \mathbb{I}\left[ f_D\left(r\left(P,a\right)\right) \leq \tau \right],
\end{equation}
where $\mathcal{J}(P,a)$ and $r\left(P,a\right)$ denotes the performance loss and residual signal caused by any attack vector $a$ on a plant $P \in \mathcal{P}$. Here $\mathbb{I}(a <b)$ denotes an indicator function that takes a value of $1$ when $a<b$ and $0$ otherwise. 

Since the adversary only knows $\mathcal{P}$, the adversary injects an attack signal to increase the risk $R_m$ by solving the following optimization problem \cite{anand2023risktac} 
\begin{equation}
R_m \; {\triangleq} \; \sup_{a}\; \underset{P \in \mathcal{P}}{\mathcal{R}} \left[ X(P,a)\right],
\end{equation}
where $\mathcal{R}$ denotes a risk measure.

The risk measure $\mathcal{R}$ selected by the adversary is closely tied to the operator's knowledge of the process dynamics. For example, consider a perfect operator as described in Figure~\ref{fig:knowledge}. If the adversary designs an attack which is stealthy for the nominal plant, the perfect operator can easily detect such an attack. In this case, the worst-case stealthy attack against a perfect operator is an attack that remains stealthy across all possible plant models $P \in \mathcal{P}$. On the other hand, if the attack is designed to be stealthy only with respect to the nominal model, an imperfect operator may misinterpret the resulting signals as consequences of model mismatch rather than malicious activity. 

%% file: Application.tex
\subsection{Power grids}
It was first pointed out in \cite{liu2011false} that a coordinated attack can be staged in a power grid without being detected at the Bad Data Detector (BDD). The paper also provided an undetectable attack design algorithm against power grids. Thenceforth, the security of power grids has been studied extensively.

Suppose an adversary injects false data into one of the measurement signals $i \in \mathcal{M}$, where $\mathcal{M}$ is the set of all measurements in a power grid. Then, similar to \eqref{eq:security}, \cite[(11)]{sandberg2010security} defines a security metric as the number of measurements $j \in \mathcal{M}\backslash i$ that has to be corrupted to remain stealthy. Let us denote this metric as $M_1$, which is identified to be non-convex and NP-hard \cite[Theorem 1]{hendrickx2014efficient}. However, an upper bound for the metric can be easily computed as pointed out in \cite[Section III.B.1]{dan2010stealth}. Under some assumptions, a method to recover the value of the metric $M_1$ using the upper bound was proposed in \cite[Section III.C]{dan2010stealth}. 

A metric similar to $M_1$ was studied in \cite[(11)]{kim2011strategic}. Computing the metric is identified to be NP-hard, and thus an $\ell_1$ norm-based relaxation is adopted \cite[(12)]{kim2011strategic}. The conditions to recover the value of the metric $M_1$ exactly, using the solution of the $\ell_1$ norm-based relaxed metric, was studied in \cite{sou2013exact}.
The metric $M_1$ studied in \cite{sandberg2010security} does not consider the magnitude of the attack vector. To this end, \cite[(13)]{sandberg2010security} defines a security metric that quantifies the least amount of energy required to be injected in the measurement signals $j \in \mathcal{M}\backslash i$ by the adversary to remain stealthy. Let us denote the metric by $M_2$. The value of the metric can be determined using an LP and was demonstrated on an IEEE 39 bus \cite{TEIXEIRA201111271}. The paper \cite{pan2018cyber} extends the metric proposed in \cite[(11)]{sandberg2010security} to consider both DoS attacks and data injection attacks. The authors show that the value of the metric for such joint attacks can be determined using a Mixed Integer Linear Program (MILP). 

The distributed state estimation algorithm was considered in \cite{vukovic2013security}. Then, the security metric was defined similarly to $M_1$. For different attack strategies (see \cite[Section 3.2]{vukovic2013security}), an approximate solution to the non-convex security metric $M_1$ was obtained. The performance of the attack strategies was demonstrated via numerical simulations on an IEEE 118 bus. 

When the state estimates of the power network under attacks are severely biased, the operator is forced to close the loop over the corrupted state estimate. This can correspond to re-dispatching in power networks. Thus, \cite{teixeira2012optimal} defines the security index as the number of corrupted sensors needed to cause the operator to initiate a re-dispatch. In a similar line of work, \cite{teixeira2014security} defines impact index as the maximum state deviation caused.

The security of a High Voltage Direct Current (HVDC) transmission system is studied in \cite{ding2020quantifying}. Using a small-signal model, the authors model the closed-loop using a continuous-time LTI system. The adversary is modeled as an additive uncertainty \cite[(16)]{ding2020quantifying}. Security is quantified as the largest $2$-norm of the additive uncertainty block under which the closed-loop system is stable. If the plant dynamics are unknown, \cite{rojas2012analyzing} proposes a data-driven method to estimate the plant’s $H_{\infty}$ norm. The security metric in \cite{ding2020quantifying} is then given by the inverse of the $H_{\infty}$ gain (small-gain theorem).

\subsection{Other applications}
Before we conclude this section, we aim to give readers practical examples widely used in the literature to study the security of NCS. By doing so, we hope to provide a foundation of common examples that researchers can use to compare the effectiveness of different security metrics, helping to create a starting point for a benchmark example for studying security in NCSs.

For instance, the set-based impact metrics were used to study the security of cooperative adaptive cruise control \cite{huisman2023impact, yang2023risk}, automated vehicles \cite{yang2023attack}, cooperative driving \cite{huisman2024optimal}. The reachable set of an adversary in the finite horizon was demonstrated on a two-stage water distribution plant in \cite{morris2017design}. The set-based impact metrics were also studied to study the security of a water distribution network \cite{ahmed2017model}, a chemical plant \cite{tunga2018tuning,milovsevic2018quantifying}, and a chemical plant with a heat exchanger  \cite{murguia2019model} which is a benchmark in fault detection literature \cite{watanabe1983fault}.

The impact metrics for aging attacks described in \cite{escudero2019prevention} were demonstrated on a DC motor \cite{pantonial2012real,escudero2020aging}. The resource-based metric proposed in \cite{milovsevic2020actuator} was demonstrated in an IEEE-14 bus model \cite{kodsi2003modeling}. 

The performance degradation caused by stealthy attacks, characterized by cyclic output-to-output gain \eqref{eq:COOG}, was used to study the security of a power generating system with a hydro turbine \cite{anand2023risktac}. The power generating system was also used to depict the threat of zero-dynamics attack in \cite{park2019stealthy}, and covert attacks in \cite{gallo2024switching}. The metric output-to-output gain \eqref{eq:OOG} was used to study the security of the Swedish power transmission grid in \cite{nguyen2025security}, and the IEEE 14-bus network in \cite{nguyen2023optimal}.

Instead of developing metrics for generic LTI systems, the work \cite{taormina2017characterizing} studies the effect of nine different attack types on water distribution networks. The effects of the attacks are quantified using three different metrics: (a) total tank overflow, (b) total time at a low level, and (c) relative variation in the pumps’ total power consumption. The effect of the attack was studied on a benchmark water distribution network \cite{ostfeld2012battle}. Similarly, the effect of the resilience metrics proposed in \cite{segovia2020cyber} was studied using the Tennessee Eastman (TE) control challenge problem \cite{ricker1993model}.

%% file: Mitigate.tex
Until now, we discussed methods to quantify security using (a) the impact caused by adversaries and (b) the resources needed by the adversary. Once security is quantified, the next step is to design a secure NCS. In this section, we recall the works mentioned in the previous sections and discuss the mitigation strategies associated with each security metric.  A brief summary of the various works reviewed, the metric they use, and their mitigation strategies are presented in Table~\ref{tab:impact-mitigation}. 
\begin{table*}[ht]
\caption{Summary of impact metrics and corresponding mitigation techniques.}
\label{tab:impact-mitigation}
\centering
\begin{tabular}{|p{4cm}|p{4.5cm}|p{6.5cm}|}
\hline
\textbf{Paper Reference} & \textbf{Impact Metric} & \textbf{Mitigation Technique} \\
\hline
\cite{murguia2020security,murguia2017reachable,hashemi2022co} & Volume of reachable states & Controller/detector design.\\
\hline
\cite{anand2020joint} & OOG in \eqref{eq:OOG} & Controller/detector design.\\
\hline
\cite{anand2022riskacc} & OOG for uncertain systems & Controller design.\\
\hline
\cite{pessim2020state} & Closed-loop stability & Gain-scheduled controller design.\\
\hline
\cite{muller2018risk} & $H_2$ norm \eqref{eq:Jez:H2} & FIR controller design.\\
\hline
\cite{feng2017resilient,zhang2019resilient} & Stability duration under DoS & Impulsive controller design.\\
\hline
\cite{lin2023secondary,lin2023secondarysos} & State safety under attacks & Secondary controller design.\\
\hline
\cite{kafash2018constraining,kafash2018constrainingcdc} & Reachable set volume & Input saturation limit design. \\
\hline
\cite{escudero2019prevention, escudero2022analysis} & State safety under attacks &Input saturation limit design. \\
\hline
\cite{escudero2022enforcing,escudero2023safety} & State safety under attacks & Input filter design.\\
\hline
\cite{gallo2021design,gallo2024switching} & OOG & Multiplicative watermarking (input filtering). \\
\hline
\cite{fang2019two} & None & Two-way coding (input filtering). \\
\hline
\cite{murguia2016characterization,murguia2016cusum,umsonst2018game,umsonst2023bayesian} & Alarm threshold vs FAR & Detector tuning.\\
\hline
\cite{milovsevic2020actuator} & Resource metric & Sensor allocation. \\
\hline
\cite{milovsevic2020security} & Scenario-specific impact scores & Sensor allocation.\\
\hline
\cite{nguyen2022single, nguyen2022zero, nguyen2023optimal, nguyen2023security, nguyen2024scalable} & OOG &  Sensor allocation. \\
\hline
\cite{tang2019linear, chong2015observability, chong2020secure, fawzi2014secure, yang2018multi} & Attack detectability & Sensor attack correction. \\
\hline
\end{tabular}
\end{table*}
\subsection{Optimal controller and detector design}
When the impact is quantified as the volume of the reachable set of states by the adversary \eqref{eq:Re}, the works \cite{murguia2017reachable,murguia2020security} propose a convex algorithm to design the observer gain ($K$ in \eqref{D}) and the fault detectors ($V$ in \eqref{D}) that minimizes the impact. However, when solving such design problems for a stable plant, a trivial solution is to set the controller and detector gain to zero. Setting the feedback gain to zero guarantees that the attacks over the network do not affect the control performance. To avoid this trivial solution, \cite{hashemi2022co} proposes an approach by introducing additional constraints. 

When the impact is quantified using the OOG in \eqref{eq:OOG}, the design problem might be ill-posed for a system with unstable zeros \cite{teixeira2019optimal}. The work \cite{anand2020joint} proposes an algorithm to design the controller gain that minimizes the impact in \eqref{eq:OOG} in the presence of unstable zeros. The framework was extended to uncertain systems (information structure $(1,2)$ in Figure~\ref{fig:knowledge}) in \cite{anand2022riskacc}. 
The work \cite{wang2020optimal} considers a metric similar to the OOG in the finite horizon. For a prescribed value of the metric (say $\beta$), the work proposes an attack design algorithm such that the impact is {not} smaller than $\beta$. 

The work \cite{pessim2020state} considers a linear parameter varying system under DoS attacks. Then, the authors propose a gain-scheduling state-feedback convex controller design algorithm to stabilize the closed-loop system.

The work \cite{muller2018risk} proposes a convex design algorithm against covert attacks. Here, the authors consider a set of finite impulse response (FIR) filters ({say} $\mathcal{C}$). Then, the controller design problem reduces to finding a convex combination of the predefined filter to minimize the impact of covert attacks, which is given by \eqref{eq:Jez:H2}. The design algorithm was extended from \cite{muller2017risk}.

The work \cite{feng2017resilient} proposes an algorithm to design impulsive controllers that maximize the duration for which the closed-loop system is stable under a DoS attack. Here, an impulsive controller uses dynamical observers with measurements-triggered state resetting. In the case of full-state measurement, the authors show that the impulsive controller is optimal. In the case of partial feedback, the authors quantify the optimality gap. A similar resilient control design was proposed for sampled-data control systems in \cite{zhang2019resilient}.

\subsection{Secondary controller design}
When the plant is (marginally) unstable, it would be unwise to stabilize the plant over the network. In such a scenario, consider a controller with two parts: a primary controller, which is over the network, and a secondary controller, which is free of attacks and stabilizes the plant. Thus, the control input to the plant is the sum of inputs from the primary and secondary controllers. Given a set of attack-free sensor inputs for the secondary controller, the work \cite{lin2023secondary} derives sufficient conditions that ensure the states always stay within a predefined safe set. Such conditions can be checked using convex SDP. The design algorithms are extended for non-linear systems in \cite{lin2023secondarysos,lin2025secondary}.

\subsection{Input filter design}
The work \cite{kafash2018constraining} proposes an algorithm to design the saturation limits of the actuator (similar to $\bar{u}$ and $
\underline{u}$ in \eqref{eq:Jez:inf})
to reduce the reachable set of states by the adversary defined in \eqref{eq:reachset}. Designing the saturation limits conservatively can also affect the nominal performance. The paper \cite{kafash2018constrainingcdc} develops a design tradeoff between the nominal and attack performance. 

Attacks on the actuator channels can also cause rapid aging of the physical actuators. The paper \cite{escudero2019prevention} studies the effect of aging attacks and proposes an algorithm to design the actuation limits such that the reachable set of states does not intersect the dangerous states. The design algorithm was extended for unstable systems in \cite{escudero2022analysis}. Such aging attacks can also be predicted and prevented by simulation techniques \cite{escudero2020security}. 

Another mitigation technique is to design an input filter on the plant side that filters any harmful attacks. Applying such filtering techniques, \cite{escudero2022enforcing} and \cite{escudero2023safety} guarantee that the plant states do not leave the safe set for actuator and sensor attacks, respectively. Consider an NCS equipped with a $\chi^2$ detector and an adversary injecting stealthy sensor attacks. Then \cite{escudero2023safetyifac} constructs a set of all possible inputs $u[k]$ such that the states of the plant always lie in the safe set $\mathcal{S}_x$ defined in \eqref{eq:safeset}. In other words, let us define the set 
\begin{equation}\label{eq:input:set}
\mathcal{R}_{u,k} = \left\{ u[k] \in \mathbb{R}^{q} \Big| a_u=0, (\ref{P})-(\ref{eq:det:logic}), \eqref{eq:safeset}
\right\}
\end{equation}

The paper \cite{escudero2023safetyifac} proposes a convex SDP to determine an ellipsoidal outer approximation of the set $\mathcal{R}_{u,k}$. The plant is considered safe if the input generated by the controller (which is affected by the attack signal) lies in the outer approximation of the set $\mathcal{R}_{u,k}$. 

An extension to the filter design approach in \cite{escudero2022enforcing} is to design dynamic filters at both the plant and the controller. Such a technique is called multiplicative watermarking \cite{gallo2021design} and does not suffer from performance losses in the presence of attacks (see Figure~\ref{fig:covert:mwm}). In a nutshell, multiplicative watermarking assigns a dynamic filter, say $\mathcal{F}_1$, at the sensor output and assigns a filter with the dynamics $\mathcal{F}_1^{-1}$ at the actuator. The filters are designed such that when the adversary does not precisely know the filter dynamics, the value of the worst-case performance loss in \eqref{eq:OOG} caused is minimized. To obscure the dynamics of the filters from the adversary, a switching multiplicative watermarking was proposed in \cite{gallo2024switching}. The two-way coding algorithm proposed in \cite{fang2019two} is a special case of multiplicative watermarking.

\subsection{Detector tuning}\label{sec:norm:mit:tune}
The attack impact is a function of the alarm threshold $\tau$ \cite{umsonst2018game}. To detail, when the detector threshold is low, the attacker causes low impact, but the operator suffers a high FAR. On the other hand, when the threshold is high, the attacker causes a high impact, and the operator has a lower FAR. Clearly, there is a relation between attack impact and FAR \cite{anand2025feasibility}. In \cite{murguia2016characterization}, the authors provide a constructive approach to design the detector threshold that satisfies the FAR when each sensor measurement is endowed with a CUSUM detector. The work was extended to a vector case in \cite{murguia2016cusum}. In \cite{umsonst2023bayesian}, the authors propose a Bayesian game-theoretic approach to design the threshold that minimizes the impact. 

\subsection{Sensor/actuator placement}
It was shown in \cite{milovsevic2020actuator} that placing additional sensors, albeit unprotected, can increase the resource metric. Since a high value of resource metric can indicate reduced attack likelihood, the paper proposes a polynomial time algorithm to (sub)optimally place the sensors to increase the value of resource metric. 

If an operator is interested in allocating protected sensors/actuators without solving for the impact, as it can be computationally intensive, then \cite{milovsevic2020security} proposes an algorithm where each scenario is associated with a predefined value of impact. That is, in the case of zero-dynamics attacks, \cite[Table 1]{milovsevic2020security} proposes an impact value of $0/1/2/3$ depending on the feasibility of the attack. Once such a map for impact is defined, the allocation problem can be solved in polynomial time. 

For multi-agent systems, where the individual agents are modeled by a single integrator dynamics, the sensor allocation algorithm was studied in \cite{nguyen2022single}. Here, the objective was to allocate the sensors to minimize the impact in \eqref{eq:OOG}. The work was extended to uncertain systems \cite{nguyen2022zero} and double integrator dynamics \cite{nguyen2023optimal}. The double integrator dynamics was applied to study the security of an IEEE-14 model. Finally, the computation complexity of the allocation algorithm was addressed in \cite{nguyen2023security}  and \cite{nguyen2024scalable} using the concept of dominating sets \cite[Definition 3]{nguyen2023security} and positivity \cite{farina2000positive}, {respectively}. 

\subsection{Attack correction}
Let us consider that some of the sensors are under attack. The work \cite{tang2018sensor} establishes the conditions under which the attack-free sensor output can be reconstructed. The attack correction algorithms were discussed in \cite[Section 7]{tang2019linear}. Other related attack correction algorithms can be found in \cite{chong2015observability,chong2020secure,fawzi2014secure,yang2018multi}.

%% file: Future.tex
\subsection{Scalable metrics}
The security metrics are usually determined by solving an SDP, except for very few works \cite{milovsevic2017exploiting,hirzallah2018computation,anand2024scalable}. An SDP has at least a cubic time complexity in the size of the system state \cite{boyd2004convex}. Thus, numerical solvers can easily determine the SDP-based metrics for small-scale systems. However, the methods do not scale well for large-scale systems \cite{nguyen2023security,arnstrom2025efficiently} such as the Swedish power grid, which has approximately $3100$ states \cite{thorslund2017swedish}. Thus, scalable metrics are missing from the literature. An initial promising direction is to explore the impact on positive systems since they are shown to possess scalable properties in general \cite{rantzer2018tutorial}. 
\subsection{Metrics for nonlinear systems}
The security metrics in the literature are usually developed for a linear system, and the works that consider nonlinear systems do not consider any stealthiness constraints\cite{khazraei2022resiliency,wang2017security}. Thus, there is a lack of metrics to quantify the security of nonlinear systems against stealthy attacks. Although addressing this issue can be challenging, restricting the class of nonlinearities (as incrementally exponentially stable nonlinear systems \cite{khazraei2022resiliency}, parameter varying systems \cite{pessim2020state}) or the class of attacks (denial-of-service attacks \cite{de2016networked,kato2021security}, zero-dynamics attack \cite{park2018stealthiness}), or explicitly designing the attack policy \cite{nguyen2024optimal} can be a promising research direction. A few related works on nonlinear system security can be found in \cite{shang2024nonlinear} and references therein.
\subsection{Metrics for data-driven control algorithms}
The data-driven control (DDC) paradigm has recently gained attention in the control community \cite{de2019formulas,martin2023guarantees}. In short, DDC aims to design a control algorithm without any knowledge of the model. There are several challenges in studying the security of DDC. Firstly, since most of the literature considers a model-based detector, a challenge lies in developing a data-driven detection algorithm \cite{krishnan2020data,zhao2022data,shinohara2025detection}. Secondly, developing data-driven detectors might be more challenging when no attack-free data is available \cite{anand2023data,yan2025secure}. Thirdly, if the operator is implementing a DDC (as it has no process knowledge), how do we define the knowledge/policy of the adversary \cite{thapliyal2023data}? Do we consider a model-based attacker to analyze the worst-case or a data-driven attacker? Providing insights into these research questions can be impactful to the research community. Some related works on security of DDC are briefly presented next. The work \cite{osama2020robust} develops a robust learning framework when some portions of the data are corrupted. The work \cite{russo2021data,russo2021poisoning,arnstrom2024data} develops a data-driven attack design strategy, whereas the works \cite{khazraei2022resiliency_L4DC,anand2025analysis} design an attack policy against data-driven controllers.
\subsection{Realistic knowledge setups} 
As mentioned before in Section~\ref{sec:risk}, most of the literature considers a problem setup where the adversary and the operator have access to accurate process knowledge, which is an unrealistic assumption. Since most of the literature assumes a worst-case setup, the mitigation strategies can yield poor nominal performance. Thus, characterizing the impact of other knowledge setups discussed in Figure~\ref{fig:layout} and developing corresponding mitigation strategies is an open research direction.
\subsection{Integrated attack-resilient control framework}
The control network is usually affected by issues such as interference, delays \cite{tipsuwan2003control}, packet dropouts \cite{wu2007design}, bandwidth limitations,  etc. The literature presents various algorithms to design controllers that are resilient to these issues \cite{zhang2001stability}. Attack-resilient control strategies that are also resilient to such network issues are lacking in the literature. Similarly, the IT literature has developed various methods to mitigate network attacks. Developing mitigation strategies by combining ideas from IT security and control-theoretic security, such as encrypted control \cite{darup2021encrypted}, is an interesting research avenue.  
\subsection{Impact of attack combinations}
In general, the security of data injection attack is studied in majority of the literature. The security of other classes of attacks, such as replay attacks, DoS attacks, and routing attacks, is not well-studied. Additionally, it is not necessary for an adversary to conduct only one class of attacks. For instance, an adversary can perform a DoS attack on some sensor channels and inject false data in other channels \cite{pan2018cyber, li2025secure}. Quantifying the security of generic attack classes and combinations of attacks is an open research venue.
\subsection{Design for privacy and security}
The notions of security and privacy are usually considered separately. In general, an NCS is defined as secure when there are minimal performance losses in the presence of attacks. Similarly, an NCS is defined as private if an adversary is unable to learn the internal dynamics of the NCS using the data transmitted over the network. Thus, it is logical to expect that security and privacy issues are related. For instance, when there is a privacy breach, the adversary can develop an approximate plant model based on the disclosed data. Based on the model, the adversary can construct \emph{stealthy} attack signals that go unnoticed by the operator, resulting in a security breach. Although intuition presents that security and privacy issues are related, many of the works in the literature consider them as standalone problems. Thus, developing mitigation strategies by considering both privacy and security is an open research avenue. Some initial results were discussed in \cite{mukherjee2021secure,abdalmoaty2023privacy}.